\documentclass[utf8]{FrontiersinHarvard} 

\usepackage{url,hyperref,lineno,microtype,subcaption}
\usepackage[onehalfspacing]{setspace}

\def\keyFont{\fontsize{8}{11}\helveticabold }
\def\firstAuthorLast{Weaver {et~al.}} 
\def\Authors{Kimberly A. Weaver\,$^{1,*}$, Jenna M. Cann\,$^{1,2,3}$, Ryan Pfeifle\,$^{1,3,4}$, Miranda McCarthy\,$^{1,5}$, Laura D. Vega\,$^{3,6,7}$, Ron Gamble\,$^{1,3,7}$, Teresa Monsue\,$^{3,7,8}$, Kyla Mullaney\,$^{1,3,9}$, Mainak Singha\,$^{3,8,10}$, Erini Lambrides\,$^{1,3,7}$, Jeffrey McKaig\,$^{1,4}$, Isabella Carlton$^{11}$, Kelly Whalen\,$^{1,4}$, Emma Kleiner\,$^{5}$, Atul Mohan\,$^{8,12}$, Subhajeet Karmakar\,$^{4,6}$, Ann Hornschemeier-Cardiff\,$^{1}$, Herbert Ortiz\,$^{13}$, Claudio Ricci\,$^{14}$, Lynne Valencic$^{1,15}$, Brandon Coleman\,$^{1,4}$, Kaylee DeGennaro$^{16}$, and Ruchi Pandey$^{1,15}$}
\def\Address{$^{1}$X-ray Astrophysics Laboratory, NASA Goddard Space Flight Center, Greenbelt, MD 20771, USA\\
$^{2}$Center for Space Science and Technology, University of Maryland, Baltimore County, 1000 Hilltop Circle, Baltimore, MD 21250, USA\\
$^{3}$ Center for Research and Exploration in Space Science and Technology II (CRESST II), NASA/GSFC, Greenbelt, MD 20771, USA\\
$^{4}$ Oak Ridge Associated Universities (ORAU), 100 Orau Way, Oak Ridge, TN 37830, U.S.\\
$^{5}$ City University of New York, 205 East 42 St, New York, New York 10017, USA\\
$^{6}$ Exoplanets and Stellar Astrophysics Laboratory, NASA Goddard Space Flight Center, Greenbelt, MD 20771, USA\\
$^{7}$ University of Maryland, College Park, 3972 Campus Dr. College Park, MD 20742, USA\\
$^{8}$ Catholic University of America, 620 Michigan Ave. NE, Washington, DC 20064, USA\\
$^{9}$ Southeastern Universities Research Association (SURA), 1201 New York Avenue NW, Suite 430, Washington, DC 20005, USA\\
$^{10}$ Astroparticle Physics Laboratory, NASA Goddard Space Flight Center, Greenbelt, MD 20771, USA\\
$^{11}$ University of Utah, 201 Presidents Circle, Salt Lake City, UT 84112, USA\\
$^{12}$ Solar Physics Laboratory, NASA Goddard Space Flight Center, Greenbelt, MD, USA\\
$^{13}$ New York University, Department of Physics, 726 Broadway, Room 1005, New York, NY 10003, USA\\
$^{14}$ Department of Astronomy, University of Geneva, ch. d'Ecogia 16, 1290, Versoix, Switzerland\\
$^{15}$ Johns Hopkins University, 3400 N. Charles St., Baltimore, MD 21218, USA\\
$^{16}$ Department of Physics, Brown University,182 Hope Street, Providence, RI 02912, USA
}
\def\corrAuthor{Kimberly Weaver}

\def\corrEmail{Kimberly.A.Weaver@nasa.gov}

\begin{document}
\onecolumn
\firstpage{1}

\newcommand{\prl}{Physical Review Letters}

\newcommand{\apjl}{Astrophysical Journal Letters}

\newcommand{\aap}{Astronomy \& Astrophysics}

\newcommand{\aapr}{Astronomy \& Astrophysics Review}

\newcommand{\apj}{Astrophysical Journal}

\newcommand{\apjs}{Astrophysical Journal Supplement}

\newcommand{\mnras}{Monthly Notices of the Royal Astronomical Society}

\newcommand{\nat}{Nature}

\newcommand{\ssr}{Space Science Reviews}

\newcommand{\nar}{New Astronomy Reviews}

\newcommand{\aj}{Astronomical Journal}

\newcommand{\pasa}{Publications of the Astronomical Society of Australia}

\newcommand{\solphys}{Solar Physics}

\newcommand{\araa}{Annual Review of Astronomy and Astrophysics}

\title[Ultra High Resolution X-ray Imaging]{The Need for Ultra High Resolution X-ray Imaging} 

\author[\firstAuthorLast ]{\Authors} 
\address{} 
\correspondance{} 

\extraAuth{}

\maketitle

\newpage

\begin{abstract}

 This paper discusses the broad science case for obtaining milliarcsecond to microarcsecond astronomical imaging resolution in the soft to medium-energy X-ray band ($\sim0.5$ to $\sim8$ keV). Astronomy across much of the electromagnetic spectrum has been fundamentally transformed with a rapid increase in ground-based and space-based capabilities to examine celestial objects on small scales that relate directly to their relevant physical processes. X-ray imaging capabilities, however, have fallen far behind observations at longer wavelengths. As such, without decisive advances in X-ray imaging, we will be unable to uncover key phenomena on the smallest astrophysical scales, leaving entire classes of high-energy discoveries beyond our reach. Here we describe several science goals for which high quality X-ray imaging is crucial and the status of some current technologies or mission concepts that would be required for these advances. In particular, we discuss the Accretion Explorer, a mission architecture under current study for a dispersed aperture X-ray interferometer.

\tiny
 \keyFont{ \section{Keywords:} X-ray emission, Astrophysics, AGN - Active Galactic Nucleus, black holes, Stellar physics, Interferometry, jets, Dual AGN} 
 
\end{abstract}

\section{Introduction}

Today's most fundamental astrophysical questions are presented in the National Academies Decadal Survey -- {\it Pathways to Discovery in Astronomy and Astrophysics for the 2020s} \citep[hereafter Astro2020;][]{national2021decadal}. The highlighted themes of Cosmic Ecosystems and New Messengers and New Physics call out priority science areas of Unveiling the Hidden Drivers of Galaxy Growth and opening New Windows on the Dynamic Universe. Key astrophysical questions posed within these areas are: (1) {\it How do gas, metals, and dust flow into, through, and out of galaxies?}, (2) {\it How do supermassive black holes form?}, and (3) {\it How is their growth coupled to the evolution of their host galaxies?} A third major science theme that emerges in Astro2020 is Worlds and Suns in Context, which emphasizes the value of investigating our Sun and solar system as a foundation for the search for habitable exoplanets. The challenge of answering the questions in all of these science focus areas requires us to examine celestial objects on spatial scales relevant to their astrophysical nature and their intrinsic physical processes. These regions can be microscopic on the sky, ranging from powerful stellar flares on other stars to enigmatic accretion disks around supermassive black holes (SMBH).

The public has become familiar with the detailed and beautiful images that are revealed at long electromagnatic wavelengths with missions such as the Hubble Space Telescope (HST), the James Webb Space Telescope (JWST), the W. M. Keck Observatory, the Very Large Array (VLA), the Atacama Large Millimeter / Submillimeter Array (ALMA), and now the Event Horizon Telescope (EHT). Taken together, these telescopes provide resolutions that range from sizes of 50 milli-arcseconds (mas) to 13 micro-arcseconds ($\mu$as) on the sky. Fifty mas can resolve physical sizes of protoplanetary disks within the Milky Way, while 13 $\mu$as resolves the ``shadow" of the Milky Way's central black hole at about 52 million km across, or a size that fits inside the solar orbit of Mercury. These telescopes allow scientists to separate the light from multiple, distinct regions, which can provide a sometimes significant advantage over examining only the integrated light from a much larger area on the sky.

 At shorter wavelengths, or higher photon energies, the state-of-the-art imaging provided with X-ray telescopes is 0.5 to 1 arcsec (as), which relates to much larger integrated sizes of a few hundred astronomical units within the Milky Way or a few parsecs (pc) on the sky for the nearest galaxies. Specific to the studies of SMBH, such resolution is insufficient for a detailed study of their innermost processes or to answer key questions posed by Astro2020. Accreting SMBH are strong sources of X-rays, and they are surrounded by gas and dust where X-ray and UV photons can be scattered, absorbed, and/or reflected. Hard X-rays associated with most active galactic nuclei (AGN) are thought to originate from photons scattered by a hot corona. To date, we can only guess as to the physical make-up of this region, with different models lending credence to competing theories of accretion. The suggested coronal geometries range from a point source located above the SMBH to a `sandwich' model that sits above and below the accretion disk \citep{bambi2021}. The corona appears to be located within a few tens of gravitational radii of the SMBH (e.g., \citet{dai2010, kara2016}), but a lack of spatially resolved imaging prevents us from directly confirming its nature. For even the nearest X-ray bright SMBH, M87, (D $\sim20$ Mpc, at $\sim6$ billion solar masses; \citet{Lu2023}), the gravitational radius ($r_{\rm g}$) would be $\sim19$ billion km ($6\times10^{-4}$ pc across, $\sim4\times10^{-6}$ as) and $\sim$ five to six orders of magnitude below our current resolution limits.

A current limitation of the field of exoplanets is that, as long as stars are treated as point sources, any surface features of the star manifest themselves merely as systematic errors in the measured exoplanet parameters \citep{Czesla2009,Uttley2021}.
For exoplanets that may be, like Earth, located in the ``habitable zone" of their parent stars, higher resolution X-ray images would allow us to discern fine structures and dynamic processes in stellar environments that are currently blurred or entirely invisible. We would gain unprecedented insights into stellar magnetic activity, flare morphologies, and atmospheric escape processes, which are vital for understanding the impact of the host star on planet habitability, atmospheres, and planetary formation. Spatially resolved studies of stars are therefore critical for a better understanding of the exoplanets they host.

With the wealth of discoveries that remain beyond our grasp in high-energy astrophysics, it is crucial to advance our technologies to probe the smallest angular scales. In this paper, we evaluate the potential of ultra high-resolution X-ray imaging to achieve transformative scientific discoveries with a telescope of significantly better angular resolution compared to the current state-of-the-art of $\sim$$0.5''$--$1''$ available with Chandra, or currently proposed for future missions that use traditional X-ray optics. To compare with the two most recent NASA X-ray concepts that have undergone studies, this includes the expected resolutions of AXIS \citep[$1.5''$; ][]{koss2025} and Lynx \citep[$0.5''$; ][]{gaskin2019}. All of the science areas we discuss would significantly address the goals of Astro2020 and respond to stated needs of the astronomical community.

\section{The Current Landscape of X-ray Observational Capabilities}

\begin{figure}[h!]
\begin{center}
\includegraphics[width=14cm]{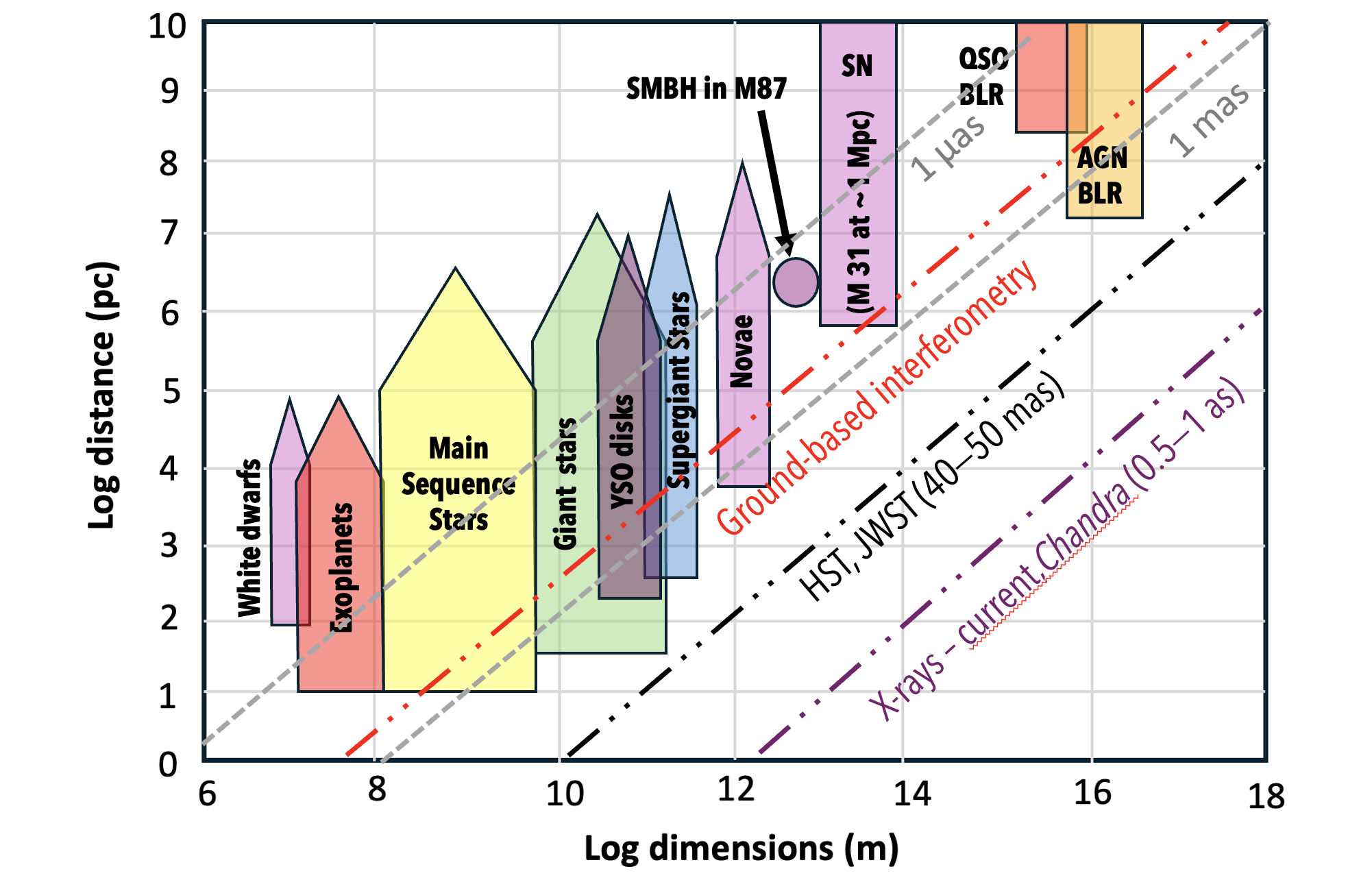}
\end{center}
\caption{\textbf{Multi-wavelength mismatch.} Current X-ray imaging angular resolution falls behind the radio, IR and optical bands, across the range of astronomical objects in distance-size parameter space, preventing synergistic multi-wavelength studies of a wide range of astronomical objects at comparable spatial scales. Ground-based telescopes now routinely operate in the mas regime (red dot-dashed line here). Color coding for boxes is based on the red to violet optical spectrum. Objects dominated by longer wavelength radiation are red; X-ray dominated objects are violet.  All of these physical scales are inaccessible to current X-ray capabilities (purple dot-dash line). Figure adapted from \citet{Rinehart2019}}
\label{fig:sizes}
\end{figure}

X-ray photons originate from some of the most powerful and explosive events in the universe.\ Foundational discoveries about black holes (BHs), BH acccretion, stellar explosions, jet physics, and stellar coronal mass ejections (CMEs) will result from increasing X-ray imaging angular resolution closer to what has been achieved at longer wavelengths (Figure \ref{fig:sizes}). There is still an enormous amount of information that remains hidden at or below the mas level for typical distances to different groups of astronomical objects relative to their intrinsic size, and the bulk of astronomical objects require telescope technology to reach mas levels and beyond (Figure \ref{fig:sizes}).\ 

This does not imply that X-ray astronomy has stalled. In contrast, within the past $\sim25$ years since the launch of the Chandra X-ray Observatory and the X-ray Multi-Mirror Mission (XMM-Newton), there has been a flood of discoveries with these and other X-ray telescopes like the Neil Gehrels Swift Observatory (Swift), the Nuclear Spectroscopic Telescope Array (NuSTAR), the extended ROentgen Survey with an Imaging Telescope Array (eROSITA), and now the Imaging X-ray Polarimetry Explorer (IXPE) and the X-ray Imaging and Spectroscopy Mission (XRISM). These findings include progress on black hole and stellar feedback, galaxy evolution, galaxy mergers, AGN, jet physics, galaxy outflows, planetary nebulae, supernova remnants (SNRs), young stellar objects (YSOs), and exoplanet environments. Just a look at the Chandra photo album\footnote{\href{chandra.harvard.edu/photo/}{chandra.harvard.edu/photo/}} or the XMM-Newton image gallery\footnote{\href{www.cosmos.esa.int/web/xmm-newton/image-gallery}{www.cosmos.esa.int/web/xmm-newton/image-gallery}} provides a sense of the breadth of this work.

Within this 25-year time frame, there has also been a dramatic increase in non-imaging X-ray capabilities. XRISM, for example \citep{10.1117/1.JATIS.11.4.042026}, provides a thirty times improvement in spectral resolution compared with Chandra \citep{2002PASP..114....1W} and XMM-Newton\citep{2001A&A...365L...1J}. IXPE provides crucial X-ray polarimetry information \citep{2025AJ....170..325F}. NuSTAR provides significant sensitivity and coverage of the hard X-ray spectrum beyond 12 keV to $\sim80$ keV \citep{2013ApJ...770..103H}. The Neutron Star Interior Composition Exlorer (NICER) has provided extremely sensitive timing analysis opportunities \citep{2016SPIE.9905E..1HG}. However, during this same era, X-ray imaging capabilities have remained stagnant. While some new ideas have been proposed, they have been stifled by limited technological development and a lack of selected mission concepts.

We will discuss here the need for zooming into and mapping high-energy astrophysical systems on small scales down to a $\mu$as, which would directly complement the advances in spectral resolution and X-ray sensitivity and enhance the science returm of high-resolution images at other wavelengths. It is clear that this capability offers new discovery space. X-ray astronomy is long overdue in building upon state-of-the-art results achieved by multi-wavelength high angular resolution imaging.

\section{The Science}

\begin{figure}[h!]
\begin{center}
\includegraphics[width=14cm]{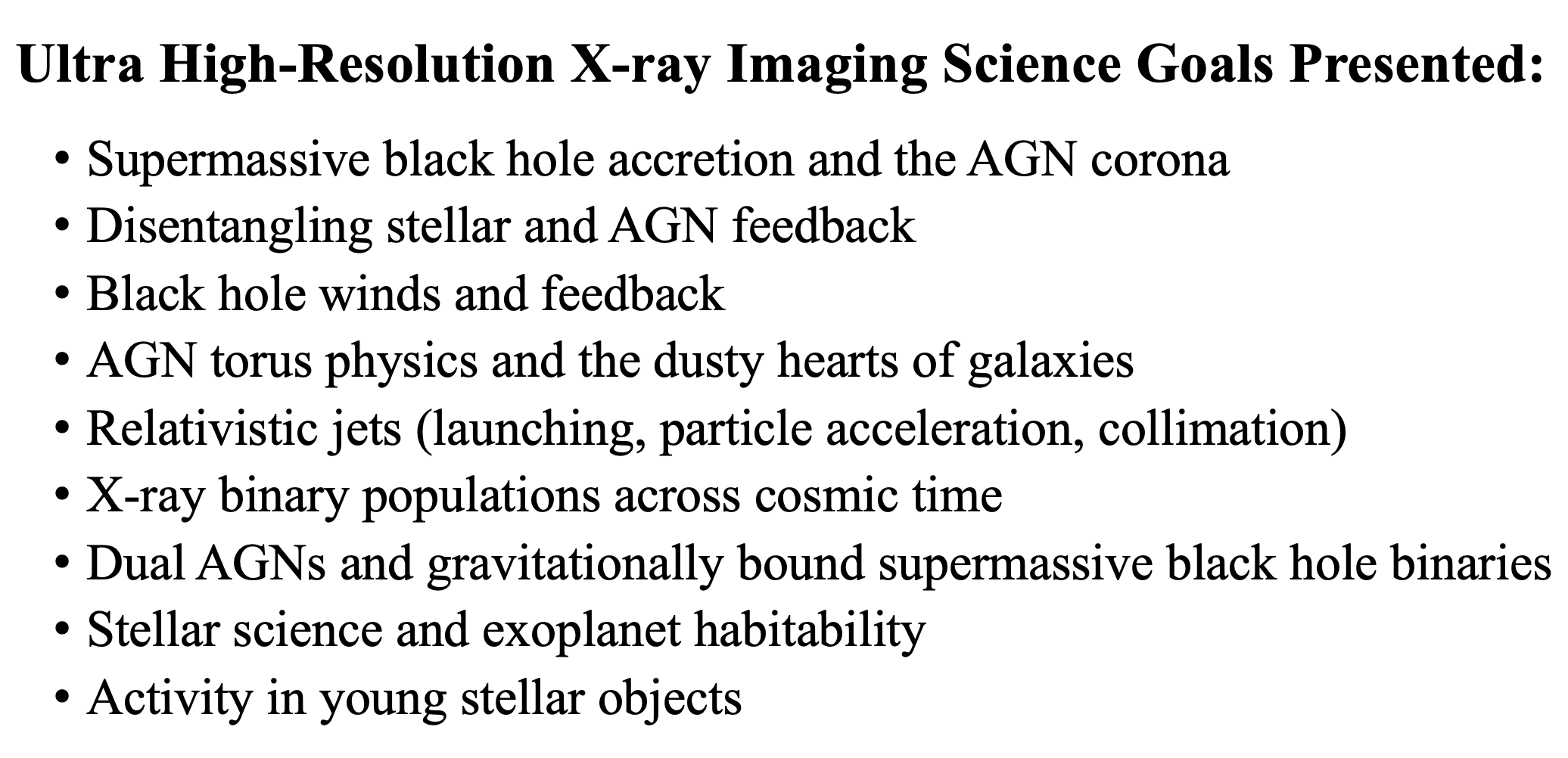}
\end{center}
\caption{Science discussed in this paper.}
\label{fig:bullets}
\end{figure}

In this section, we examine how astronomy will benefit from discrete increases in imaging resolution, stepping from $\sim0.1$ arcseconds to $\sim1$ mas, to $\sim1$ $\mu$as. The reason for discussing multiple angular scales is that different technologies are required to achieve each range of observational capability. Because the argument for science gains between different spatial resolutions is not always clear and can depend on source brightness, we mix the discussion within each science subsection. Our goal in this work is not to advocate for any single capability; rather, our aim is to demonstrate the need for this capability in general, and thereby also demonstrate the need for the development of the enabling technologies for ultra high-resolution X-ray imaging. Our motivating science was selected from a set of categories within the Astro2020 priorities with a focus on the properties of AGN, jets, galaxy mergers, X-ray binary (XRB) populations, and stellar science (Figure \ref{fig:bullets}). The Astro2020 specific science priority that would be addressed is called out in each subsection heading below.

\subsection{Active Galactic Nuclei (AGN) (Astro2020 Drivers of Galaxy Growth and Dynamic Universe)}

Accreting SMBH at the cores of AGN shine brightly in X-rays, and the small scale emission regions that impact their host galaxies on larger scales are still places that we cannot see directly in X-rays. With gains in resolution to just 0.01-0.1 arcsec, we could reach less than pc-size scales in nearby AGN to study how SMBH and their galaxies interact. A bigger challenge is presented by the shadow of a SMBH event horizon being only a miniscule $\sim4\times10^{-6}$ arcsec across even for the closest X-ray luminous SMBH, M87 \citep{Lu2023}. Even though current and next-generation observatories at longer wavelengths can reach scales of $\sim0.5$ mas (e.g., ngVLA) to $\sim$50 mas (e.g., JWST), many questions remain such as: (1) {\it How do AGN accretion disks maintain themselves?}, (2) {\it What is the relationship between the accreting system and the circumnuclear material?}, (3) {\it What are the true geometries of the inner regions of different types of AGN?}, and (4) {\it What physical structure(s) produce the complex X-ray absorption seen in many AGNs?} Constructing emission maps at different X-ray energies could probe what differentiates starburst-driven and AGN-driven outflows near galaxy centers, trace geometries of locations like the obscuring torus and the X-ray broad line region (BLR), and test what mechanisms are responsible for AGN feedback \citep[see reviews by ][]{alexander2025,kara2025}.

\subsubsection{Supermassive black hole accretion and the AGN corona} 

\begin{figure}
\centering
\includegraphics[trim={0.1cm 0.28cm 0.25cm 0.06cm},clip,width=0.4\textwidth]{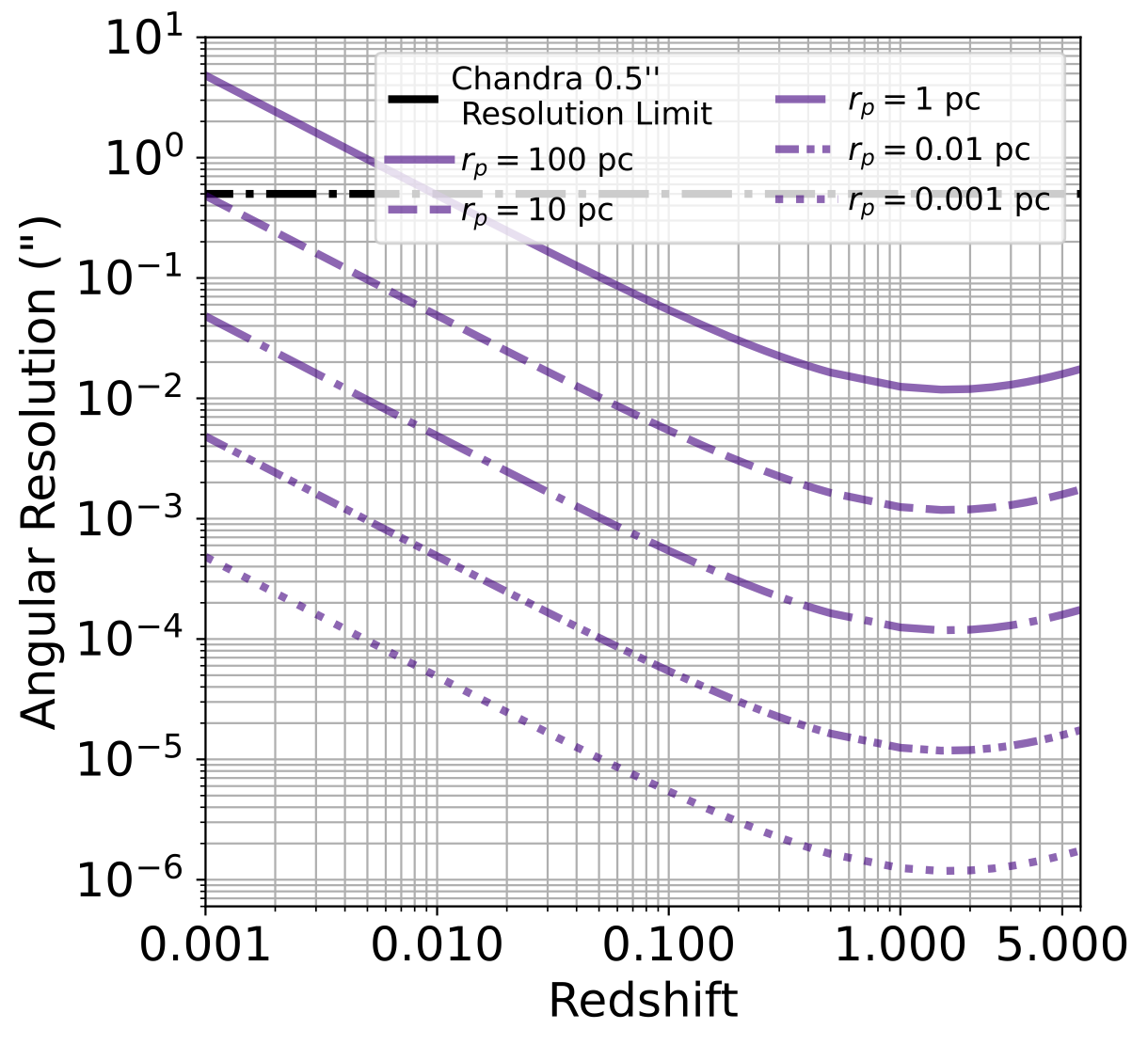}
 \includegraphics[width=0.55\textwidth]{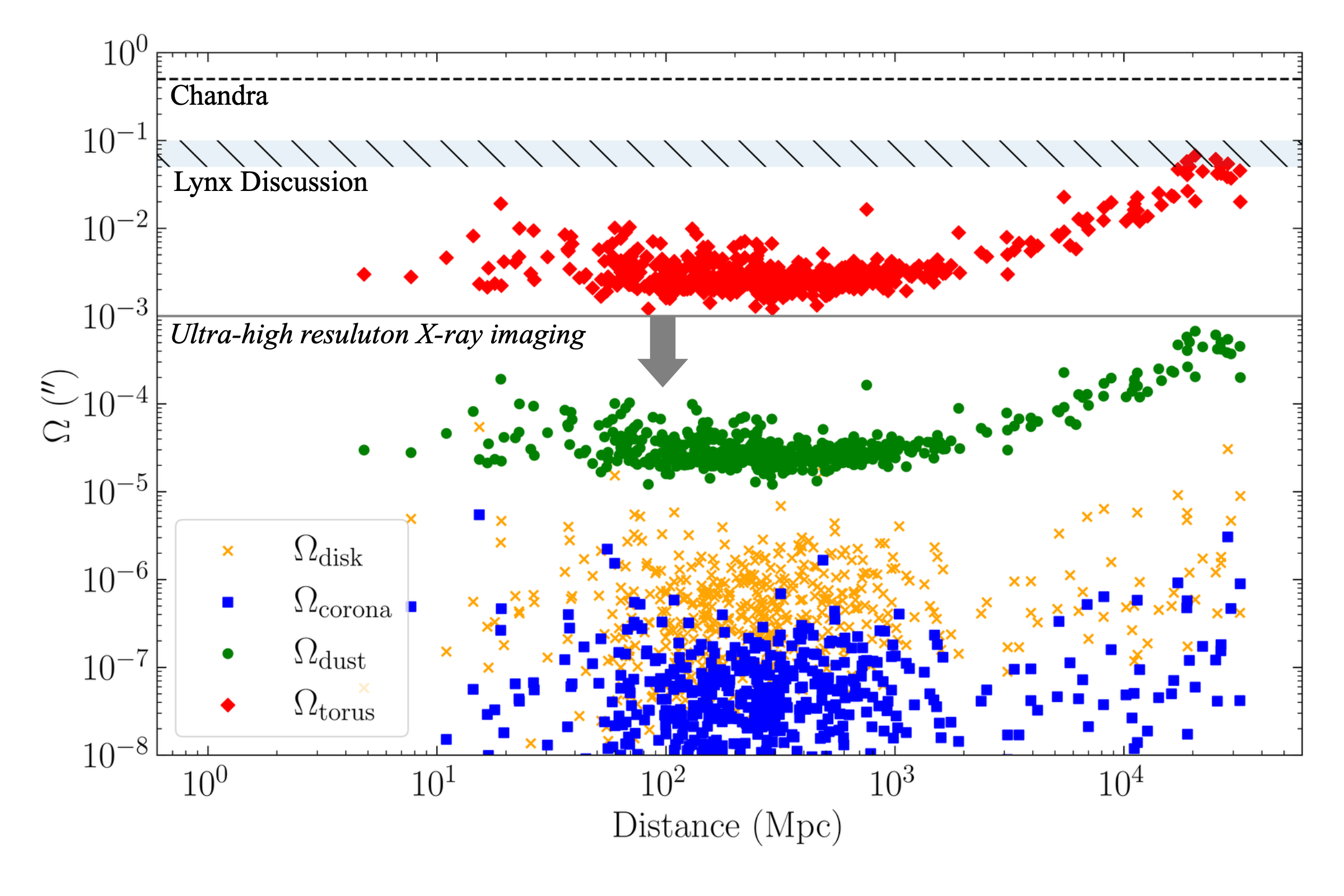}
\caption{
\textbf{Spatially resolved studies of AGN accretion are inaccessible for a wide range of spatial scales with current instrumentation.} Left: Here we plot angular resolutions needed to probe spatial scales of 100\,pc down to $10^{-3}$\,pc in the nearby and distant Universe. Right: The angular scales that could be probed for targets in the Swift-BAT AGN sample \citep{bassDR2}
as a function of distance. The blue, orange, green and red points correspond to the resolutions needed to resolve the X-ray corona, outer accretion disk, the dust sublimation radius, and the outer torus, respectively, for each target. The mas through $\mu$as angular resolutions discussed by this white paper reside under the horizontal grey line indicated by the downward arrow. As can be seen, mas resolution imaging may be sufficient to probe the dust sublimation radii of a few targets. However, an X-ray interferometer with \textit{$\mu$as resolution} is needed to be able to probe the dust sublimation radii of the entire sample, and would also open up a window to observe the outer edge of the accretion disk, and even the X-ray corona in a handful of sources.}
\label{Fig:resolution}
\end{figure}

We are increasingly learning more about the physics of the innermost regions of AGN. Without access to imaging resolutions to probe these regions directly, several indirect methods have been utilized, such as spectral energy distribution (SED) analysis to measure black hole mass \citep{gliozzi2024}, X-ray polarization measurements to assess the corona \citep{2023MNRAS.523.4468G}, X-ray reverberation mapping using time lags to measure sizes of the accretion disk \citep{2021iSci...24j2557C}, and X-ray spectral analysis to probe the inner accretion flow \citep{piotrowska2024}. But with the first ``image" of a BH from the Event Horizon Telescope, there has been significant interest in further understanding even the smallest-scale processes and structures. In fact, several current and proposed radio observatories are reaching angular resolutions able to probe AGN at pc and sub-pc scales \citep[e.g.,][]{godfrey2012,fischer2021}. However, there has been limited advancement into the high-resolution studies of complementary X-ray observations, significantly curtailing our understanding of these regions, and subsequently our understanding of the fundamentals of accretion power that can dramatically impact AGN host galaxies at all scales. We have calculated the angular resolutions needed to probe scales from 100\,pc down to $10^{-3}$\,pc as a function of redshift (Figure \ref{Fig:resolution}, left). Specifically, we have also examined the spatial scales needed to directly probe key regions of an AGN, such as the X-ray corona, accretion disk, dust sublimation radius, and the torus, as a function of redshift (Figure \ref{Fig:resolution}, right). These physical scales are not accessible with Chandra.

Despite the remarkable development in BH accretion theory over the last five decades, we lack requisite resolution to spatially resolve the accretion flow. While deep Chandra observations have revealed the structure of accreting gas in local galaxies \citep[e.g., M84; ][]{Bambic2023}, Chandra's angular resolution cannot spatially resolve any turbulent gas motion, which begins to manifest within the inner 100 pc \citep{hopkins2016}. Highly resolved X-ray images of this region would provide direct insight into the physics of accretion flows and their effects on AGN structure. The majority of AGN likely accrete through a geometrically thin and optically thick disk. Observing even a small sample of these disks accreting across a range of Eddington ratios could help confirm traditional theory or probe deviations from a thin disk structure. Deviations include environments that are gas-rich. These can facilitate BH accretion through mechanisms like Bondi accretion \citep{Bondi1952}, where the accretion flow is advection dominated \citep{Narayan1998}, quiescent and hot ($>$$10^{7}$ K). However, turbulence may complicate the process, leading to an alternate theory of chaotic accretion \citep{Gaspari2013}, in which accreted gas is rather cold ($10^{4}-10^{7}$ K) but its motion is significantly driven by turbulence. 

\begin{figure}[ht]
\centering
\includegraphics[width=1.0\textwidth]{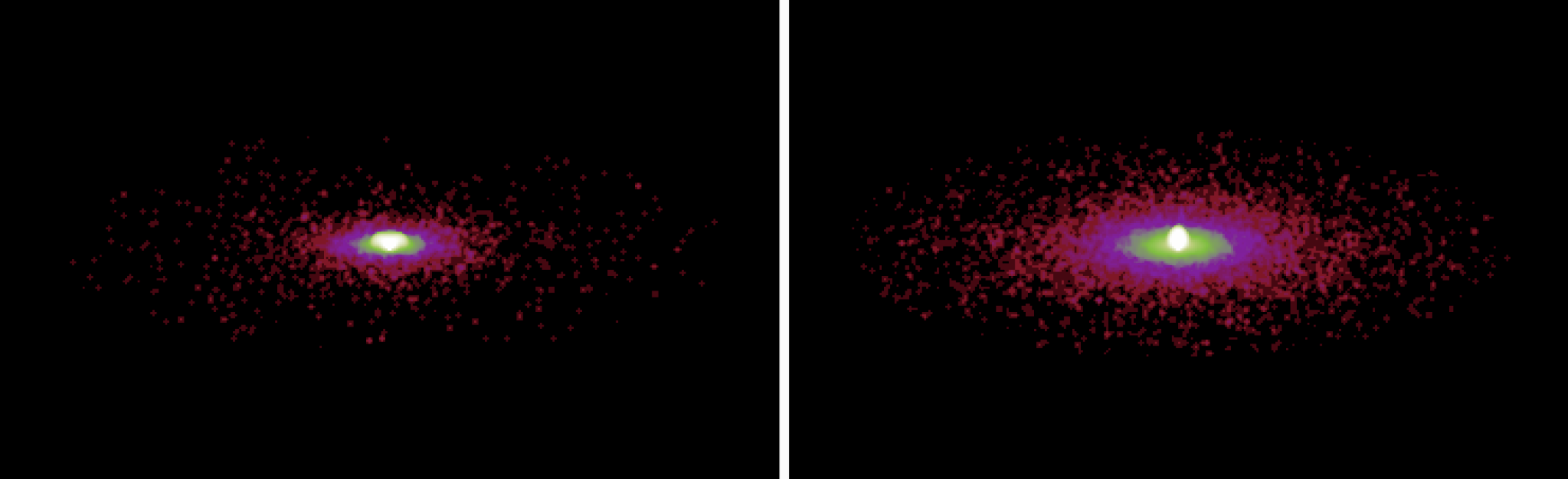}
\caption{\textbf{Simulated images of an AGN accretion disk and hard X-ray corona observed with micro-arcsecond resolution.} These simulated disks were made assuming an accreting $10^{8}\mathrm{M}_{\odot}$ SMBH at a distance of 5\,Mpc from Earth, consistent with the closest objects where we can spatially resolve the corona with $\mu$as resolution (Figure\,\ref{Fig:resolution}). Left: Accretion disk and corona system where the corona is in the form of a torus surrounding the SMBH. The torus has an inner (cross-sectional) radius of $10r_{\rm g}$. Right: Same as Left, but with a spherical corona with radius of $10r_{\rm g}$. Images were made using the X-ray ray-tracing code \textsc{RefleX} \citep{reflex2017}.}
\label{Fig:corona_models}
\end{figure}

The hard X-ray photon emission associated with most AGN is believed to originate from disk photons Compton scattering off thermal electrons in the so-called hot corona. The nature of this coronal X-ray emission is hotly debated, with different models lending credence to competing theories of AGN accretion. Over the past two decades, X-ray reflection spectroscopy and reverberation mapping studies have suggested that the X-ray corona
is compact and located within $\sim$$10r_\mathrm{g}$ from the BH \citep{2009Natur.459..540F,2011MNRAS.417L..98D,2014MNRAS.439L..26K,2014A&ARv..22...72U,kara2016,2017MNRAS.467.2566F}. This result has been supported by microlensing studies \citep[e.g.,][]{dai2010,chen2011}. A considerable portion of the coronal emission is reflected and reprocessed by the inner regions of the accretion disk. However, compact coronae close to the BH are energetically unstable, raising questions about their heating and energy storage, possibly through significant magnetic fields generated by, or advected through, the disk. Recent reverberation studies suggest that BH coronae may have a substantial vertical extent \citep{wilkins2016,hancock2023}, adding complexity to their nature and evolution. However, a lack of spatially resolved imaging prevents us from directly confirming the nature of the emission source and its connections to accretion, magnetic field structures, and/or jets \citep{Uttley2021}. 

With micro-arcsecond resolution, the coronae of the nearest AGNs come into view (Figure \ref{Fig:resolution}), allowing for the first direct images to constrain the geometry of the corona (Figure \ref{Fig:corona_models}). We note from the simulations shown in Figure \ref{Fig:corona_models} that the perceived extent of the accretion disk appears to change depending on the geometry of the corona. This is due to the fact that the coronal geometry will affect how photons are directed towards the disk and reflected back to the line of sight. This implies that even if the corona itself is not spatially resolved, the reflected emission from the accretion disk (which will be resolved in many more objects, see Figure \ref{Fig:resolution}) may be diagnostic of the coronal geometry.  

\subsubsection{Disentangling stellar and AGN feedback }


In recent decades, studies have shown that energetic winds from accreting BHs and supernovae (SNe) span various scales (sub-pc to kpc) and phases, from X-ray to molecular gas \citep{Fleutsch2021}. Unifying optical-radio AGN emission on small scales is \textit{impossible} without commensurate highly-resolved X-ray observations. Questions about winds persist: How are they launched? How do they interact with the ambient medium? What are their immediate effects? 

\begin{figure}[ht]
\centering
\includegraphics[width=0.24\linewidth]{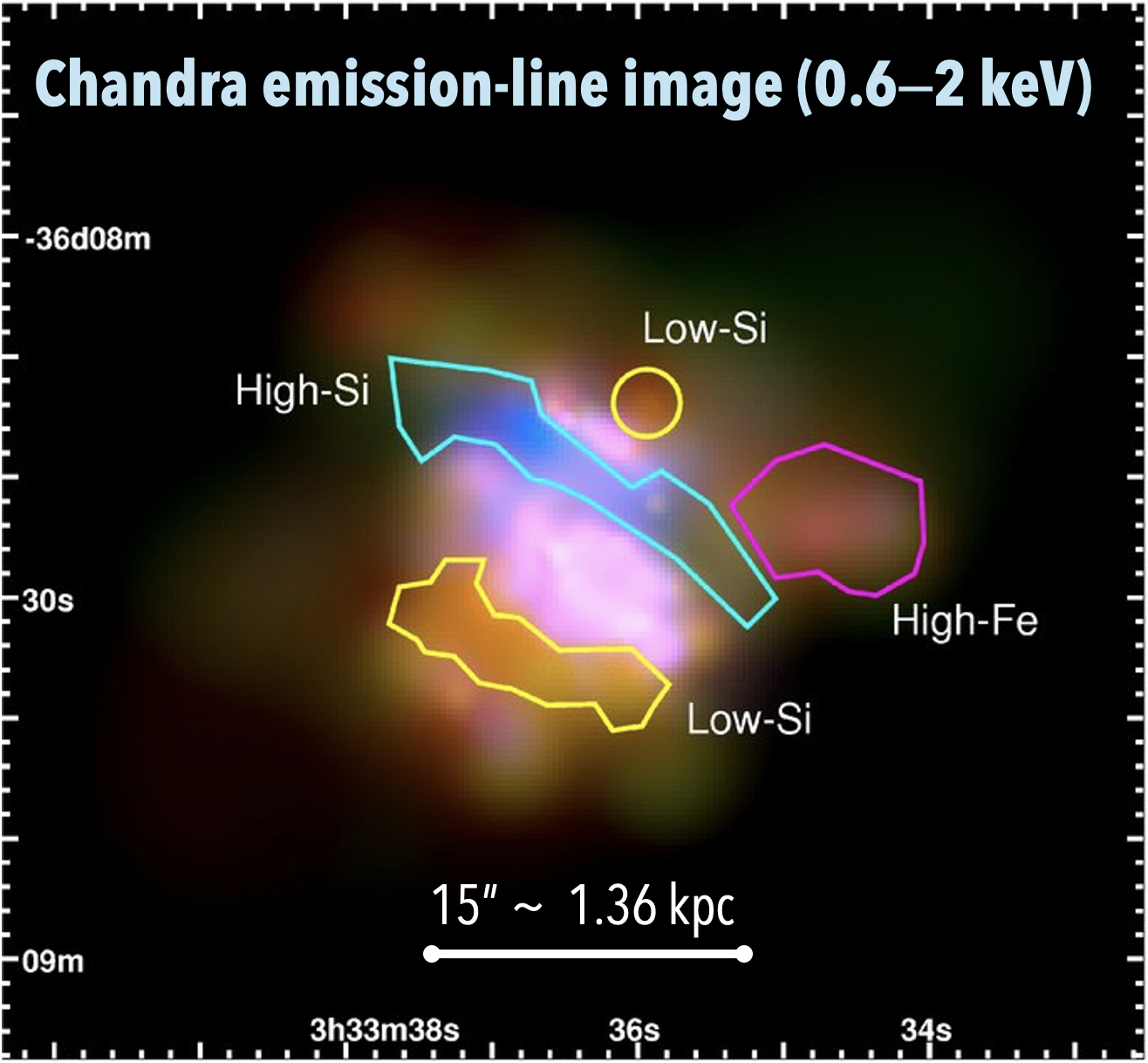}
\includegraphics[width=0.25\linewidth]{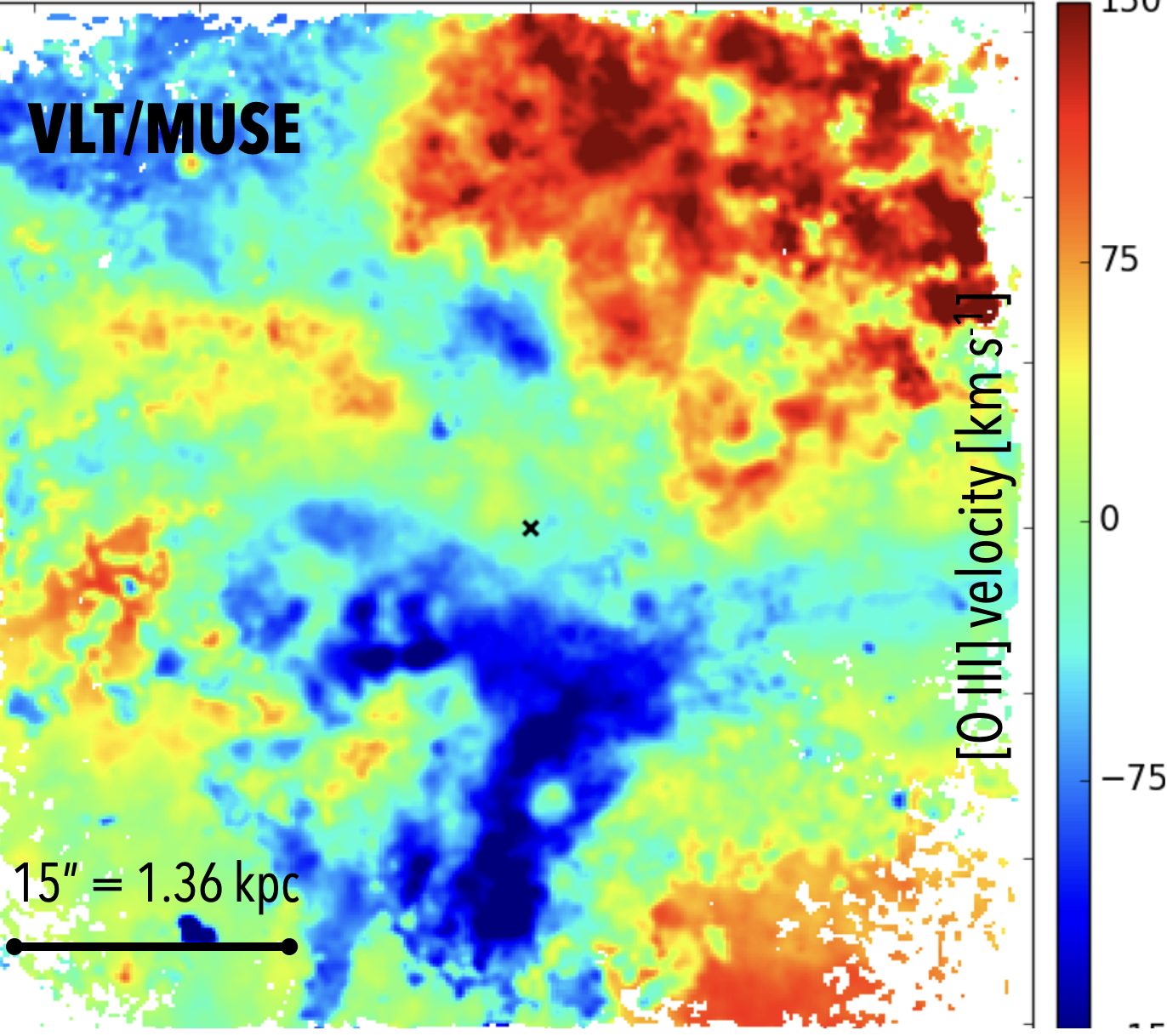}
\includegraphics[width=0.24\linewidth]{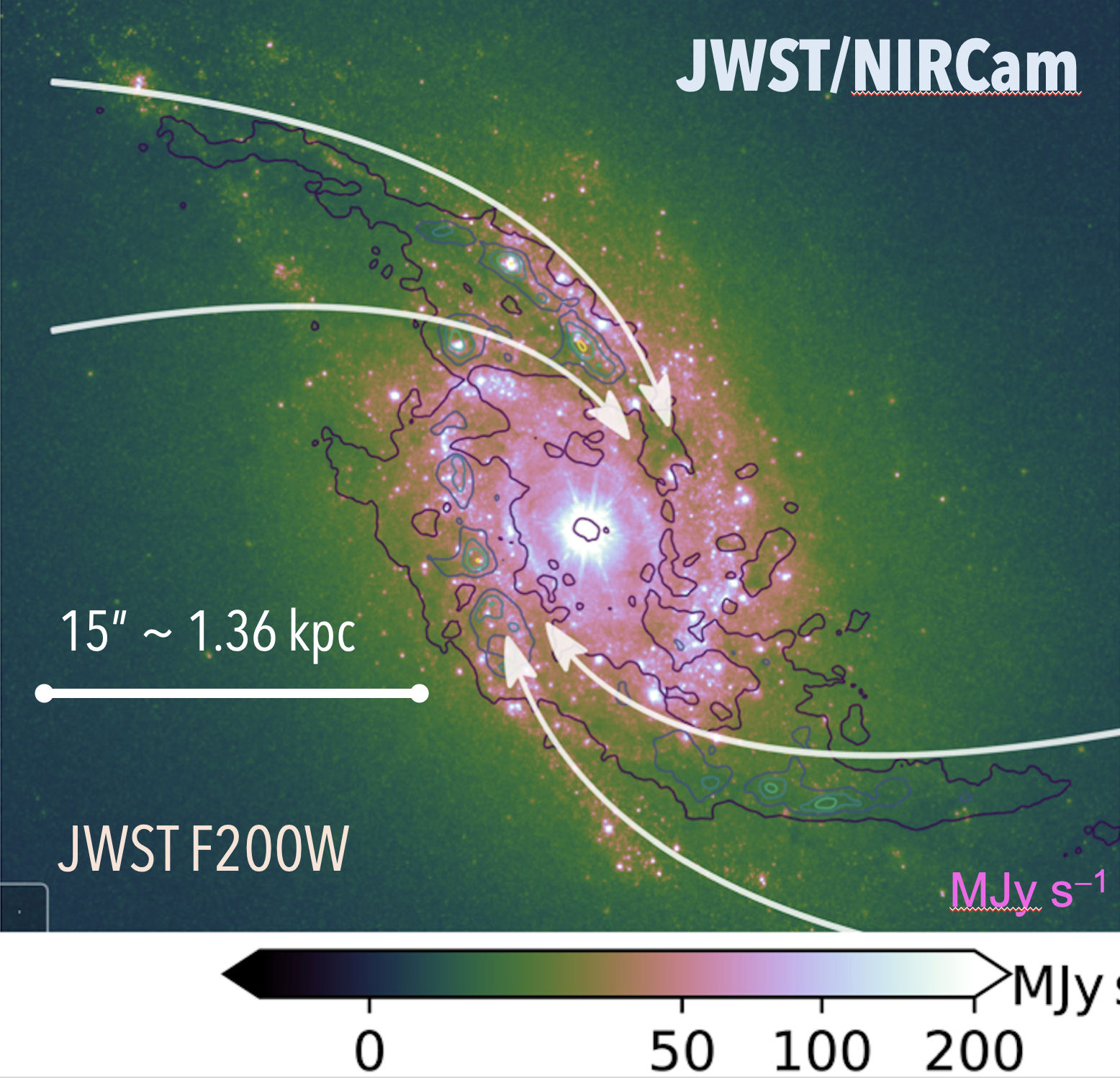}
\includegraphics[width=0.25\linewidth]{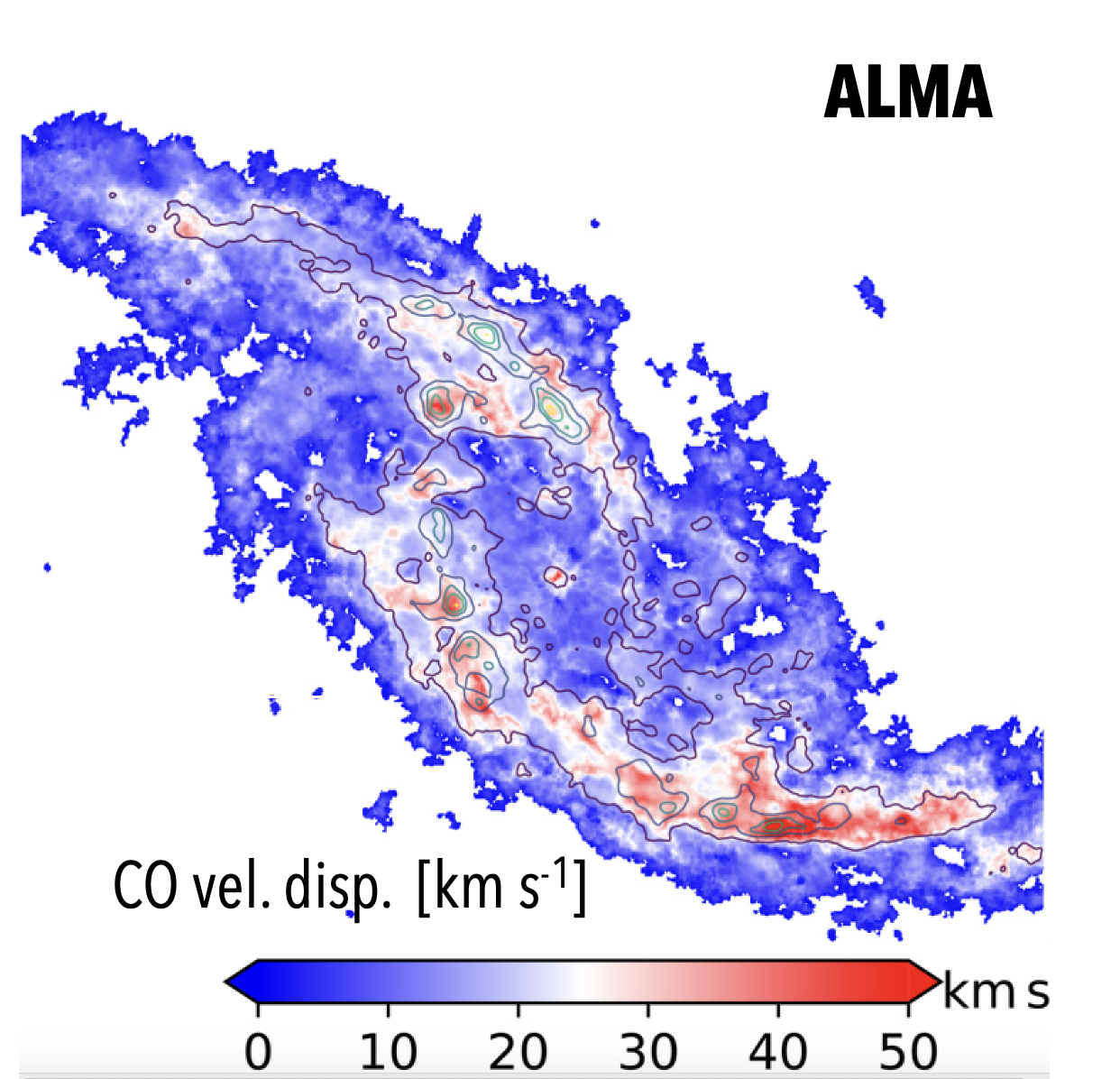}
\caption{\textbf{High spatial resolution X-ray imaging is crucial for the dissecting regions dominated by starburst/AGN feedback.}\ This figure shows different views of the Seyfert galaxy NGC 1365.\ The Chandra emission-line image (A) resolves scales of $\sim90$ pc on the sky (\citet{Wang_2009}, reproduced with permission).\ MUSE observations (\citet{Venturi2018}, reproduced with permission) (B) show shock-like structures related to the AGN driven outflow in the map of [O~III] line-of-sight velocities minus the stellar velocities. JWST NIRCam (C) and ALMA (D) resolve the central 2.5 kpc region on $<9$ pc scales, highlighting prominent starburst-driven wind signatures (\citet{Liu2023}, reproduced with permission). While Chandra shows X-ray emission spatially aligned with the AGN-driven outflow seen with VLT/MUSE, it lacks the angular resolution to resolve the potential X-ray counterparts to these finer features.}
\label{Fig:n1365}
\end{figure}

In galaxies, star formation could be regulated by the powerful winds from starbursts and supernovae \citep{Heckman2000,Heckman2002}. These winds are believed to heat up the surrounding gas and prevent the cooling of the hot circumgalactic medium \citep{Zhang2014,Haywrad2017}. In turn, the bow shocks may be located on sub-kpc scales \citep{Binder2019}. X-ray images, even with 100 mas angular resolution (possibly available with the next generation of imagers) will precisely determine these shocks and the X-ray signatures from these winds in the local universe. This will in turn determine their kinetic luminosity and quantify their role in metallicity evolution. 

Near to galaxy centers, where star formation (SF) and AGN activities interact closely, differentiating between AGN-photoionization and stellar signatures such as shocks and SNe winds is crucial to understand feedback and galaxy growth. One such case is the galaxy NGC 1365 (Figure \ref{Fig:n1365}), where soft X-rays primarily arise from thermal emission linked to SF \citep{Whewell2016}. Chandra detects lines from photoionized gas \citep{Wang_2009}, spatially aligned with the AGN-driven outflow seen with VLT/MUSE (Figure \ref{Fig:n1365}). JWST and ALMA observations reveal SNe winds on even smaller scales \citep{Liu2023}. However, current X-ray images lack sufficient angular resolution, and improving the resolution would allow us to search for X-ray structures and filaments. Achieving merely $0.1''$ 
imaging resolution would enable direct comparison of X-ray emission with optical, IR, and mm maps, confirming SNe shocks in the ring and the role of the AGN in driving the winds. 

\subsubsection{Black hole winds and feedback}
\label{windsfeedback}
\begin{figure}[ht]
\centering
\includegraphics[width=\linewidth]{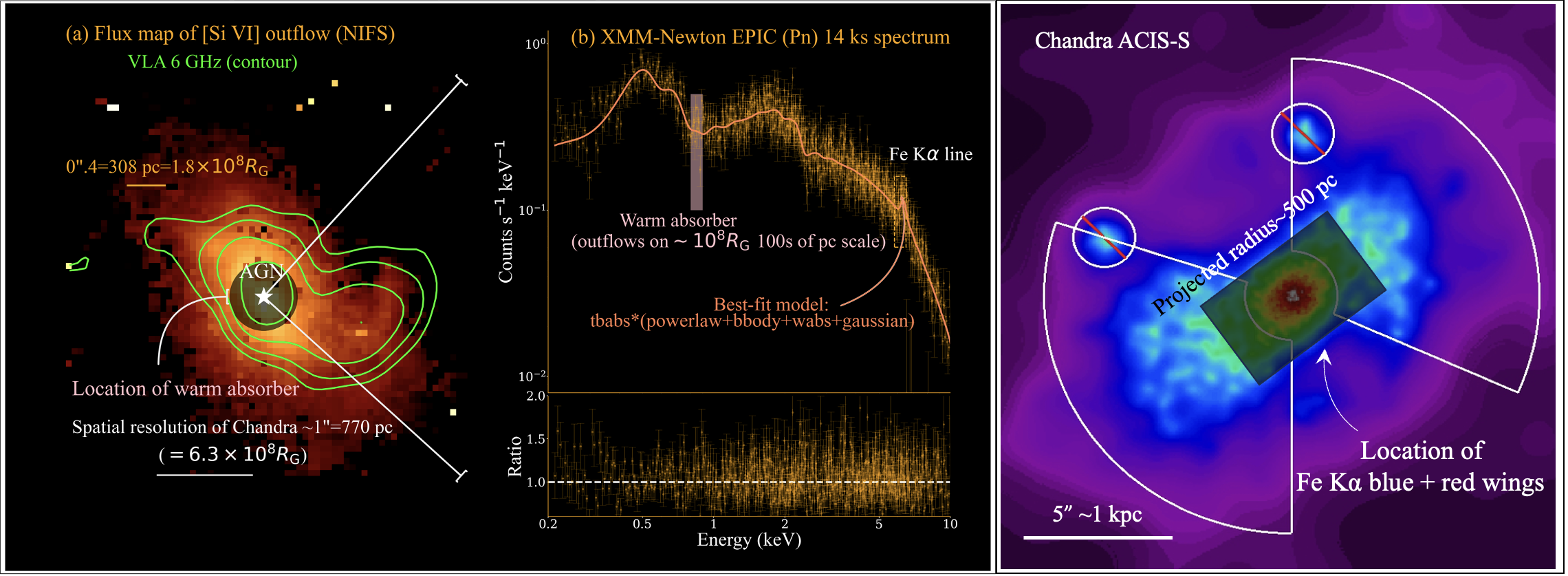}
\caption{
\textbf{An order of magnitude enhancement in angular resolution over Chandra ACIS-S would reveal the interaction of outflows with the ambient medium.} Left panel: Multi-phase outflow in nearby ($z$=0.038) radio-quiet AGN HE~0412$-$0803. Here we show archival Gemini/NIFS IFU (FWHM=$0.18''$) observations that reveal an AGN-driven outflow in [Si~VI]$\lambda1.965\mu$m, spatially overlapping with collimated radio emission seen with the VLA. The XMM-Newton spectrum shows a warm absorber feature between 0.6-0.8 keV, where photoionization modeling suggests a spatial extension of 308 pc $\sim$400 mas (gray filled circle). The angular resolution provided by current and future (planned) X-ray instrumentation 
is insufficient to spatially resolve these outflows. Right panel: UFO(v$\sim0.1c$) in the Fe-K$\alpha$ emission line in NGC~5728 (taken from \citet{TrindadeFalcao2024} and reprinted with permission). The image shown is the Chandra ACIS image from 0.3 to 8 keV with the inner region of the bi-conical outflow greyed over to emphasize that Chandra does resolve this UFO.
}
\label{Fig:outflow}
\end{figure}

The most energetic phases of accreting SMBH/AGN and XRBs accretion disk winds are spotted in X-ray wavelengths as warm absorbers and ultrafast outflows (UFOs) \citep{Tombesi2010,DelSanto2023}.\ Warm absorbers show up as absorption lines in the 0.5–2 keV band (see, for example Fig.~\ref{Fig:outflow}, left panel), often with blueshifts indicating outflow speeds of $100-2000~\mathrm{km~s}^{-1}$ \citep{laha2014}. They come from partly ionized O, Ne, and Fe, typically extending over a few hundred pc ($\sim400$ mas at z=0.04). On the flip side, UFOs track highly ionized gas moving at relativistic speeds ($v \approx 0.1-0.4 c$), seen through absorption lines of highly ionized elements like Fe XXV (6.7 keV) and Fe XXVI (6.97 keV) \citep{fukumura2015}. They are thought to originate as close as $6r_\mathrm{g}\sim7\mu$as (at z=0.04) for magnetically driven winds \citep{fukumura2014,fukumura2015}, while UV-line drivers may start from at least $60r_\mathrm{g}$ \citep{hagino2015}.\ In NGC 5728, a spatially extended (1"$\sim$500 pc) sub-relativistic outflow ($v\sim0.1c$) detected in Fe K$\alpha$ emission (see Fig.~\ref{Fig:outflow}) suggests that UFOs may also show up in X-ray emission lines and stretch over a few hundred parsecs \citep{TrindadeFalcao2024}. X-ray imaging with even a resolution of 100 mas could better resolve these extended UFOs and transform our understanding of BH winds. 

\subsubsection{AGN torus physics and the dusty hearts of galaxies}

While the standard Unified Model for AGN has withstood decades of multi-wavelength observational tests, the physical structures and their interactions remain unclear. Due to angular resolution limits, spectral and variability studies have been the only probe of X-ray emission from these inner regions, compared to direct interferometric imaging in other wavelengths \citep[e.g.,][]{jaffe2004,leftley2018,alonso-herrero2018}. Several studies paint a picture of a dynamic torus, more complex than the classical compact, static model, that includes regions of extended emission and an intricate combination of multi-phase outflowing and inflowing gas \citep[e.g.,][]{gallimore2016,alonso-herrero2019,garcia-burillo2021}.

\begin{figure}[ht]
\centering
\includegraphics[width=0.98\textwidth]{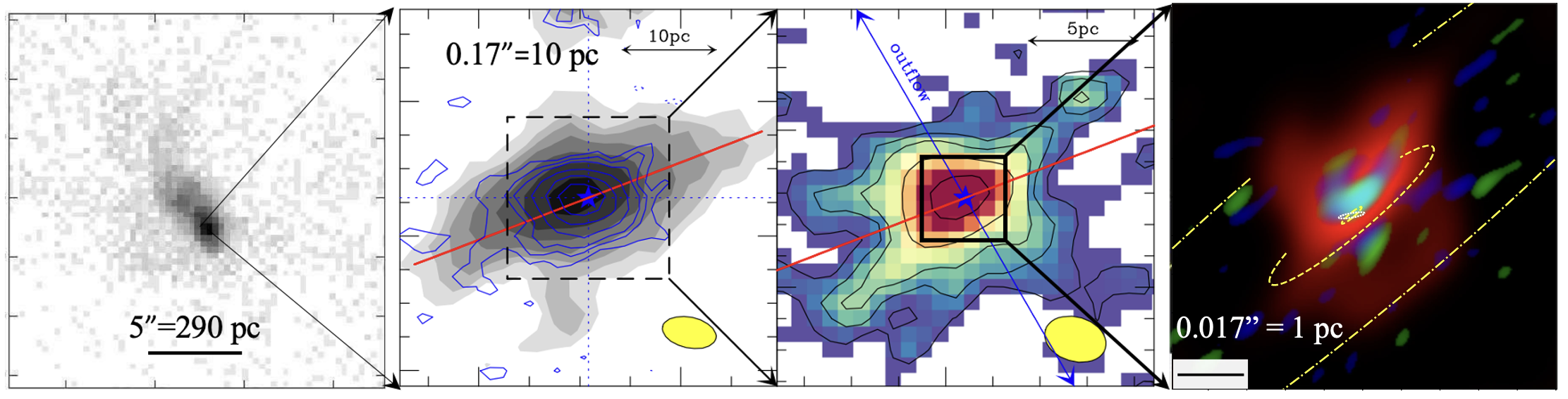}
\caption{
\textbf{
Multi-wavelength images of NGC 1068.} These are from Chandra in the $0.25 - 7.5$ keV energy band (left, \citet{Young2001}, reprinted with permission) ALMA in CO (2--1) and HCO+ (4--3) (center-left and -right;\citet{garcia-burillo2019}, reprinted with permission), and MATISSE IRBis in the IR (right; \citet{gamezrosas2022}, reprinted with permission). The angular resolutions of ALMA ($\sim$0.03$''$ -- 0.09$''$) and MATISSE (3 -- 10 mas) are significantly higher than that of Chandra and of the 0.5$''$ proposed in Astro2020. 
Having a multi-wavelength synergy with comparable resolution in the X-ray band would allow for a direct study of the nuclear obscuration of the central engine of an accreting SMBH. 
}
\label{Fig:ngc1068}
\end{figure}

Even with the $0.5''$ resolution proposed in Astro2020 \citep{national2021decadal}, a detailed imaging study of these regions will lie far beyond instrument capabilities (see Figure \ref{Fig:resolution}).\
However, the use of synergistic multi-wavelength observations with higher resolution X-ray imaging than currently available, coupled with physically well-motivated simulations, can provide deeper insight into the X-ray emitting regions. For example, the torus of the Seyfert 2 galaxy NGC 1068 has been mapped to parsec scales in the mm \citep{garcia-burillo2019} and MIR \citep{raban2009,gamezrosas2022}. MATISSE observations of NGC 1068 reveal a dusty structure on 20 mas scales ($\sim$1 pc), obscuring the central engine \citep{gamezrosas2022} 
(Figure \ref{Fig:ngc1068}).\ The structure is consistent with a torus but not in thermal and dynamical equilibrium.\ The process behind non-equilibrium physics remains unknown, and while torus-scale winds are detected in X-ray spectra, spatially resolved X-ray imaging is needed to transform our understanding of this region and its interactions with the surrounding medium. Chandra, \citep{Young2001} reveals only a roughly extended feature right across the nucleus (Figure \ref{Fig:ngc1068}). This may be related to the base of the radio jet, but at this resolution, we cannot determine what other detailed structure could exist here. 

With mas and $\mu$as resolution in the X-rays, synergistic multi-wavelength studies of the geometry and composition of the torus at similar spatial scales will become possible. There is also an intriguing possibility of observing targets which spectroscopically do not follow the unified model (e.g., obscured X-ray sources which still display broad lines). Direct imaging could help us understand the processes that are affecting the geometry of their central regions.

\subsection{Relativistic Jets and Particle Acceleration (Astro2020 Dynamic Universe)}\label{subsec:sciencecaseJets} 
Over fifty years have passed since the first thought experiments and theories about extracting energy from BHs were published. The leading concepts have centered around everything from exploiting the frame-dragging property of BHs in energy extraction \citep{Penrose1971, Williams2004, Williams1995}, to a formulation of the four laws of BH mechanics \citep{Bardeen1973}. To this day, there remain significant gaps in our understanding of the dynamic ejection of matter from BHs. Specifically, current theories lack sufficient detail on the mechanism driving this energetic extraction, and how relativistic jets are collimated in the process. Efforts to reconcile general relativistic simulations of the BH accretion mechanisms and their associated ejecta with observational signatures have only been significantly successful in post-processing schemes and codes. Higher-resolution observations are needed to inform and advance such simulations, greatly enhancing our understanding of relativistic jet creation and collimation.

As a specific example of the need for X-ray observations, radio interferometry (EHT) imaging of M87 (25 $\mu$as angular resolution) has revealed a ring-like structure with a radius of $8.4r_\mathrm{g}$ $(\sim$32$\mu$as), connected to the edge-brightened jet \citep{Lu2023}.\ The emission profile of the jet-launching region is wider than expected for a BH-driven jet, suggesting a wind associated with the accretion flow.\ Ultrafast outflows are seen in X-ray spectra but with significant uncertainty on their locations.\ High-resolution imaging of these outflows would thus be vital to constrain the location of these winds and their relationship with the jet.

\subsubsection{Jet launching} 

An open question in jet physics is their launching mechanism: whether they are powered by the extraction of rotational energy from the BH where magnetic field lines thread the horizon via the Blandford-Znajek (BZ) process \citep{BZ1977}, by cascades of high-energy particles under rotations in the Williams-Penrose (WP) mechanism \citep{Williams2004, Williams1995}, or sourced by the accretion flow itself via the Blandford-Payne mechanism (BP) \citep{BP1982}. Collectively, these processes describe the rotational energy extraction resulting in Poynting flux dominated outflows parallel to the rotation axis of the central BH. Such ejecta and initialization of particle acceleration may occur very close to the BH horizon surface within a range $5r_{g} \lesssim h_{jet} \lesssim 130r_{g}$, with $h_{jet}$ the length parameter parallel to the azimuthal coordinate. The X-ray corona is speculated to be the base of the jet \citep{markoff2005}. We have seen recently the initial emission of radio jets emerging across a two year time-span \citep{meyer2025}. With higher resolution X-ray imaging 
we can unveil the geometric relationship among the jet, disk, and X-ray corona, enabling us to directly observe the dynamic X-ray emission from jets. A focus on evolving computational solutions that compare jet launching theories mentioned above with key observables is needed. These improved simulation methodologies would include predictions tied to inferred jet and magnetic field geometries, polarization maps, and other parameters that aid in refining X-ray observations within the jet. 

\begin{figure}
\centering
\includegraphics[width=0.95\linewidth]{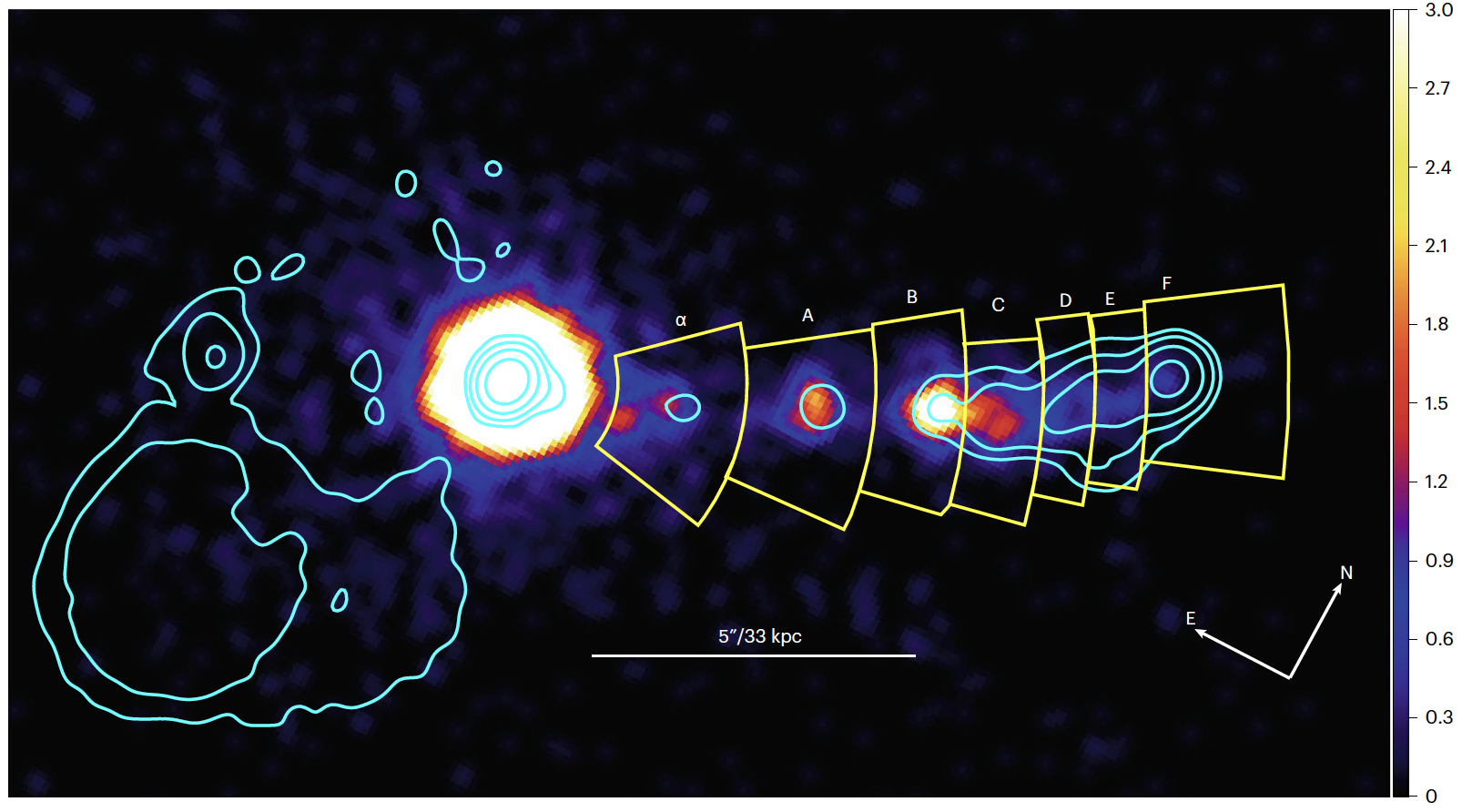}
\caption{\textbf{Chandra X-ray image of PKS 1136-135 overlayed with radio contours}. \citep{meyer2023} describes the interplay of radio and X-ray emission on a kpc scale close to the jet core. The composite image illustrates a common relationship between radio and X-ray emission within jets unified under the guise of synchrotron emission.
X-ray knots are outlined in yellow, along with the location of radio emission outlined as cyan contours.}
\label{Fig:jet}
\end{figure}

As jets often start driving shocks on smaller scales \citep{Bodo2018}, X-ray emission is expected on mas scales due to particle acceleration. However, Chandra's resolution is not high enough to confirm if particle acceleration primarily occurs on these scales where the inner radio jet is located (see Fig.~\ref{Fig:jet}). X-ray imaging with mas resolution or better will trace particle acceleration and its initialization near BHs.

\subsubsection{Particle acceleration} 
The particle acceleration mechanism in jets is not well-constrained. State-of-the-art results have been achieved in the Galactic XRB microquasar SS 433, where multi-wavelength observations have revealed hard X-ray emission at 29 pc ($\sim$$18'$), spatially overlapping with very high-energy gamma-ray emission ($\sim$$25$ TeV) \citep{Safi-Harb2022}. A possible explanation is that the innermost radio jet, moving at $\sim$$0$$.3$c, encounters a sub-relativistic shock at the inner X-ray jet, where gamma-rays result from inverse Compton scattering and X-rays from synchrotron radiation \citep{Ahronian2024}. By analyzing this synchrotron spectra from radio and X-ray emissions at more comparable resolutions, we can begin to learn in more detail the jet kinematics and subsequent magnetic field structure \citep{lister2021}. Using kinematic simulations at the length scales mentioned above, we can see what acceleration mechanisms are at play in the near-horizon zones and corona of relativistic jets. Prominent kinematic features of particle populations in jets at such high Lorentz factors ($\Gamma \gtrsim 10-100$) can provide for the predicted velocity profiles, to which a normalized distribution can be derived. 

\subsubsection{Jet collimation} 
SS 433 provides a nearby example of jet collimation. VLBA observations reveal collimated radio emission on mas scales in this eclipsing X-ray binary system \citep{2016MNRAS.461..312J}. 
While we have gained a better understanding of where jets collimate from radio studies like this, high resolution X-ray imaging will help us understand the X-ray properties in those regions, providing insight into the effects of, for example,  accretion on collimation. Recent EHT 228 GHz observations (FWHM=25$\mu$as $\sim 200r_\mathrm{g}$) have shown that in Cen A ($M_\mathrm{BH}\sim5.5\times10^{7}~M_\mathrm{\odot}$), the radio emission is collimated on incredibly small tens of $\mu$as scales \citep{Janssen2021}. However, in this case the peak radio emission appears to be shifted from its true location if it were the radio core by $\sim$$30\mu$as, leading to a possibility that the brightest points originate from shocks within the jet. The leading hypothesis for relativistic jet collimation in AGN suggests that it occurs under $200r_\mathrm{g}$ due to magnetic field pressure from disk-driven winds \citep{Boccardi2021} — a process responsible for driving UFOs (see Section \ref{windsfeedback}). Very high resolution X-ray imaging on $\mu$as scales would be required to map the regions involved. 

\subsection{XRB Populations Across Cosmic Time (Astro2020 Dynamic Universe, Drivers of Galaxy Growth)} 
\label{subsec:sciencecaseXRBs}

X-ray binaries in external galaxies and AGNs represent two of the most important extragalactic sources of X-rays and both are intimately tied to the structure, growth, and evolution of their host galaxies. X-ray emission from galaxies arises from a number of sources, with the harder (E $>2$ keV) rest-frame emission being dominated by accretion onto compact objects, either an AGN or by the aggregate X-ray emission arising from populations of accreting neutron stars (NS) and stellar mass origin black holes.  These NS and BH populations are observed as XRBs, and numerous Chandra studies have demonstrated that the X-ray luminosity functions (XLFs) and the integrated X-ray luminosities ($L_X$) of high-mass XRBs (HMXBs) and low-mass XRBs (LMXBs) scale with the star formation rates (SFR) and stellar mass ($M_{*}$), respectively, in nearby galaxies \citep{grimm2003,gilfanov2004,lehmer2019,lehmer2024}. 

There are now observational data indicating that $L_X$(HMXB)/SFR and $L_X$(LMXB)/$M_{*}$ evolve with redshift \citep{mineo2012,lehmer2016}, most notably the recent HERA measurements indicating extremely elevated X-ray emission per SFR for $z>10$ HMXB populations \citep[][see also \citet{garofali2024}]{HERA2023}. However, these relations are highly sensitive to SFR, $M_*$, metallicity, and stellar age, and observational constraints have largely depended on small, less representative samples of galaxies occupying limited host property parameter spaces. Large, more representative XRB samples of local universe galaxies ($<$200\,Mpc) have only recently become available via low angular resolution, all-sky eROSITA X-ray imaging \citep{kyritsis2024arxiv}; covering larger ranges in the aforementioned parameter spaces. New constraints derived from these samples indicate potentially significant deviations from prior results (specifically constraints on $L_X$[HMXB]/SFR) even for the local universe. 

\begin{figure*}
    \centering
    \includegraphics[width=0.48\linewidth]{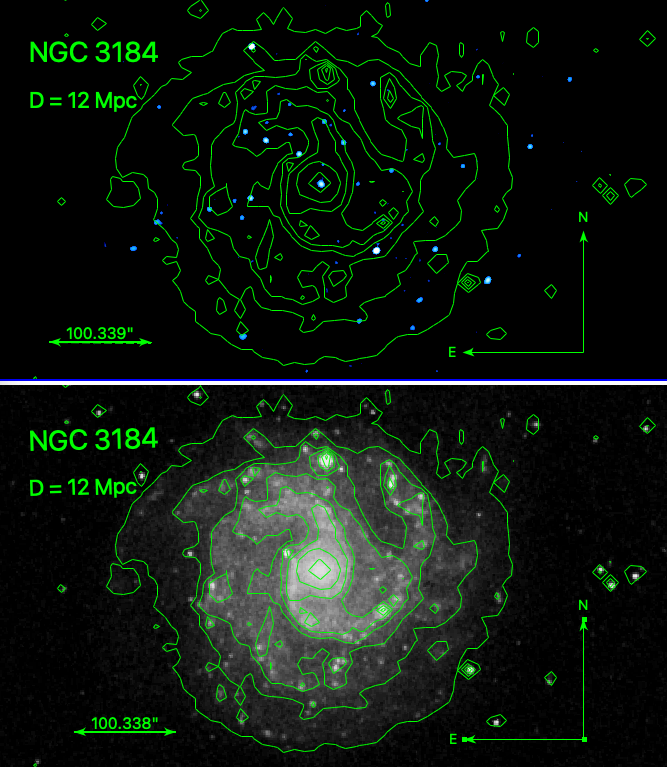}
    \includegraphics[width=0.48\linewidth]{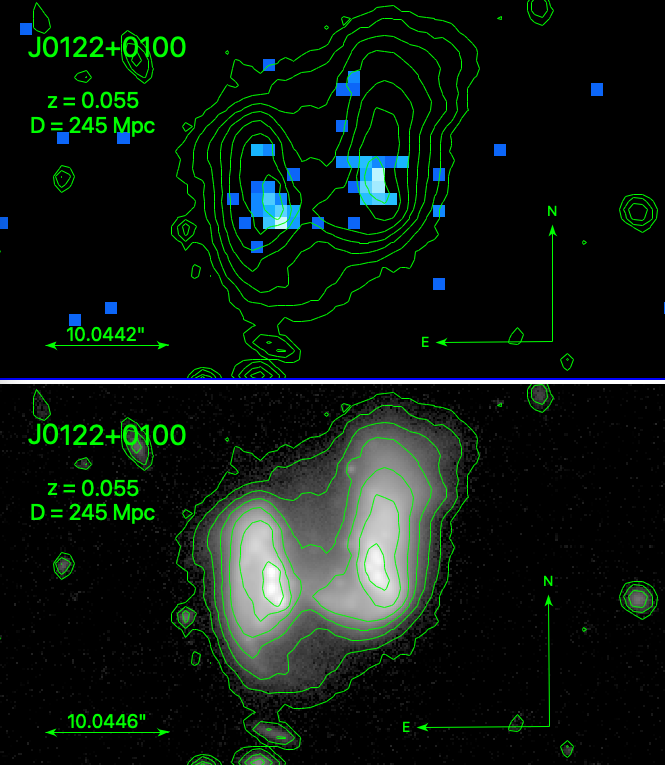}
    \vspace{+4mm}\caption{\textbf{Point source confusion with increasing redshift.} Left: Two example extragalactic environments, NGC 3184 ($D_L=12$\,Mpc) and J0122+0100 ($z=0.055$, or $D_L=245$\,Mpc). Bottom panels: optical images (DSS for NGC 3184 and Hyper-Suprime Cam for J0122+0100). Top panels: Chandra X-ray 0.3-8\,keV imaging. Distinct XRBs can be identified in NGC 3184; XRBs and AGN(s) are completely source confused in J0122+0100.} 
    \label{fig:resolvingXRBs}
\end{figure*}

Population synthesis models predict an evolution of the XRB XLFs and $L_X$(HMXB)/SFR and $L_X$(LMXB)/$M_{*}$ relations with redshift as a result of evolving host properties \citep{fragos2013}, such as metallicity and stellar age. In particular, it is expected that HMXBs will dominate the XRB emission at $z\gtrsim2.5$, and the HMXB XLF will increase by half an order of magnitude across the age of the universe \citep{fragos2013}. In fact, synthesis models and extrapolations from observations \citep{lehmer2016} predict that XRBs dominate the X-ray emission at $z>6-8$, outshining even AGNs at earlier epochs \citep{fragos2013}. However, current limitations in spatial resolution (Figure~\ref{fig:resolvingXRBs}) and sensitivity prohibit direct observational constraints on the evolution of the XLFs of HMXBs and LMXBs as a function of SFR, metallicity, stellar age, and $M_{*}$ across cosmic time. Refined XLF predictions using the latest population synthesis models \citep[e.g.][]{fragos2023,andrews2025} will play an integral role in high resolution X-ray imaging mission development.

The angular resolution limits of current X-ray facilities ($\geq$$0.5''$) impede our ability to constrain the cosmic evolution of XRBs, particularly the XLFs. Source confusion with increasing $z$ (even within $\sim$$30-50$\,Mpc) requires (a) the selection of ``normal'' (AGN-free) galaxies in statistical studies, and (b) using integrated X-ray luminosities to constrain scaling relations because distinct XRB populations cannot be resolved \cite{mineo2014,lehmer2016,kyritsis2024arxiv}. 

High angular resolution X-ray imaging that can achieve $<$$0.5''$ resolution would transform our understanding of the interplay between the XRB XLFs and host galaxy properties. Spatially-resolved X-ray detection of XRB populations, hot gas, and/or central AGNs on 100-pc scales in the local universe ($z<0.1$) and scales on the order of hundreds of pc throughout the universe could be accomplished with sensitive $\sim50$\,mas resolution imaging (see Figure~\ref{Fig:resolution}); $\sim1$~mas-scale resolution imaging with sufficient effective area would resolve spatial scales of $\leq1$\,pc in the local universe and $\sim10$\,pc scales throughout the higher redshift universe even out to $z>6$ (Figure~\ref{Fig:resolution}). Such $1-50$\,mas-scale imaging would allow for nuclear and non-nuclear X-ray source populations to be differentiated \citep[e.g., Figure 2 in ][]{hornschemeier2003} across a wide range of redshifts; in systems where the nuclear SMBH is X-ray weak, the spatially-resolved X-ray emission would be dominated by XRB-driven point sources scattered across the galaxy. Across nominal mission lifetimes, large, statistically representative samples of XRBs could be observed to place direct constraints on the XLFs in the local and distant universe. Precise constraints on how the XLFs respond to metallicity, stellar age, SFR, and host $M_{*}$ are critically important to understanding how the XLFs evolve with redshift. The populations of LMXBs and HXMBs obtained through this high resolution imaging could be used -- in concert with hard energy imaging with sensitivity and spatial resolution superior to \textit{NuSTAR} -- to identify larger populations of BH-XRBs and NS-XRBs akin to studies focusing on Chandra-NuSTAR observations of nearby galaxies \citep[e.g.,][]{vulic2018}. 

Though sensitive mas- and sub-mas-scale X-ray imaging would enable the detection of distinct XRBs within local and higher redshift galaxies \citep[e.g.,][]{hornschemeier2003,colbert2004,liu2011_pointsrc,vulic2018}, this resolution would likely come at the expense of field-of-view (FOV), meaning that even higher redshift galaxies would need to be tiled to provide sufficient spatial coverage of the XRB populations (this may be observationally prohibitive for lower redshift galaxies due to the number of required tilings). Nonetheless, even if XRB XLFs would be more observationally accessible for higher redshift (smaller angular size) galaxies, high spatial resolution mas-scale imaging would allow for X-ray imaging on par with -- or better than -- the spatial scales achievable through radio VLBI imaging. This synergy in spatial resolution between the X-rays and radio would allow for close examinations of the interplay/correlation between the X-ray and radio emission regions/properties in individual XRBs in local/lower redshift systems (or high redshift galaxies via $\mu\rm{as}$-imaging), enhancing our understanding of accretion microphysics and jet physics in these objects \citep[e.g.,][]{tomsick2003A,ribo2008}.

As the XLFs directly relate to the formation rate of XRBs and compact binaries, understanding the cosmic evolution of the XLFs will in turn inform our understanding of the high frequency stochastic gravitational wave (GW) background that arises from a large population of compact object mergers that are unresolved in the LVK band \citep[e.g.,][]{abbott2016,abbott2018_nsns_stochback}. A high spatial resolution X-ray imaging facility would come with the inevitable trade off that it could not be used for most time domain/rapid response purposes, as it would likely be associated with longer slew times and it would be temporally- and resource-prohibitive to perform large sky area localization for transient sources or gravitational wave sources/events. However, for inspiralling gravitational wave sources already localized on the sky \citep[via some combination of electromagnetic detection and/or gravitational wave detection via Advanced LIGO, Advanced VIRGO, or KAGRA; e.g.][]{abbott2017_nsns,abbott2020_prospectsforobserving}, mas- or sub-mas-scale imaging could potentially be used to study any relationship between the X-ray properties and the gravitational wave properties of individual inspiralling compact stellar mass binaries (BH-BH\citep{abbott2016}, BH-NS, and NS-NS binaries \citep{abbott2017_nsns}; see \citet{abbott2019_catalog,abbott2021_cat2,abbott2023_cat3}). Such gravitational wave follow-up observations could potentially connect pre- or post-merger accretion microphysics with the multiwavelength electromagnetic properties (optical emission, radio jets, etc.) in these compact merging systems. 

\subsection{Dual AGNs and Gravitationally Bound SMBH Binaries (Astro2020 Dynamic Universe, Drivers of Galaxy Growth)} 
\label{subsec:sciencecaseSMBHpairs} 

Decades of research have shown that galaxy mergers represent one avenue for galaxy growth and evolution as well as one potential route for the growth of the central SMBHs. Dual AGNs -- cases where two SMBHs are actively accreting in an ongoing merger of two or more galaxies -- are predicted to be observational signposts of this merger-induced SMBH growth throughout the history of the universe \citep{foord2024}. Predictions from simulations suggest dual AGNs are associated with vigorous, obscured SMBH growth, with the peak of accretion and obscuration predicted to coincide with the latest-stage of the galaxy merger process (during/after the coalescence of the host spheroids). As the host galaxies coalesce, accretion onto the two SMBHs is predicted to continue \citep[e.g.][]{blecha2018,chen2023} as they sink to the center of the host potential via dynamical friction \citep{begelman1980,amaro2023}, eventually forming a gravitationally bound SMBH binary on scales $\lesssim30$\,pc (depending upon the masses of the SMBHs). Additional angular momentum losses through viscous drag and loss cone scattering are expected to bring the binary into the continuous wave regime \cite[e.g.,][]{begelman1980,kelley2017} wherein the continued orbital decay and eventual, final inspiral and coalescence of the two SMBHs is dominated by the emission of gravitational waves \citep{begelman1980,kelley2017,burke-spolaor2018,amaro2023}. As dual AGNs are the observational forerunner to gravitationally bound binary AGNs, the accretion histories, mass evolution, and mass ratio evolution of dual AGNs directly impact observational expectations for the bound binary phase. The behavior and environments of parsec- and sub-parsec scale binary AGNs, in turn, establish the expectations for the continuous gravitational wave regime \citep{agazie2023,agazie2023} as well as the properties and timescales for the binary inspiral and coalescence \citep{amaro2023}. Dual and binary AGNs are therefore crucially important astrophysical laboratories for studying the merger-driven growth of SMBHs and the behavior, environments, and evolution of SMBH pairs in galaxy mergers, from dual AGNs during the early pericenter passages down to the inspiral of the bound SMBH binary in the gravitational wave regime. 

In the local universe ($z<0.1$), dual AGNs on scales $>10$\,kpc ($\gtrsim9''$ at $z=0.05$) can be resolved and analyzed readily with a variety of current facilities and archival resources \cite[e.g.,][]{liu2011,koss2012}. However, dual AGNs are notoriously difficult to unambiguously confirm and study at closer pair separations; as of 2020, fewer than $50$ dual AGNs with separations $<10$\,kpc had been confirmed in the literature \citep{pfeifle2025}. At separations below $1$\,kpc, the statistics are ever more scarce: to-date, fewer than $\sim 10$ dual AGNs with separations below $1$\,kpc have been robustly identified \cite[e.g.,][]{komossa2003,muller-sanchez2015,koss2023}, and only a single binary AGN has been confirmed so far \cite[separation of 7\,pc,][]{rodriguez2006}. The observational demands of spatially-resolving sub-kpc dual AGNs and pc-scale binary AGNs has left a substantial gap in our understanding of merger-driven SMBH growth and the evolution of accreting SMBH pairs at sub-kpc scales, particularly where dual AGNs transition to gravitationally bound binary SMBHs at separations $\lesssim30$\,pc.

With an on-axis angular resolution 0.5'', \textit{Chandra}'s ability to resolve closely separated dual AGNs drops off considerably with increasing redshift: while a $\sim100$\,pc dual AGN could be resolved at $z=0.01$ ($\sim43$\,Mpc), \textit{Chandra} can only resolve down to  separations $\sim200$\,pc, $\sim490$\,pc, and $\sim920$\,pc by $z=0.02$, $z=0.05$, and $z=0.1$, respectively (see Figure~\ref{Fig:resolution}). Combined with the fact that the vast majority of dual AGN candidates have been found at $z\gtrsim0.02-0.03$ (only nine dual AGNs have been confirmed at $z<0.02$, none of which have separations $<1$\,kpc, see \citealp[]{pfeifle2025}), dual AGNs with separations $\lesssim200-300$\,pc are virtually inaccessible to \textit{Chandra}; only one dual AGN with a separation of 230\,pc has been confirmed \citep[at $z=0.0347$,][]{koss2023}, but the angular separation of the AGNs ($0.32''$) falls below the angular resolution limit of Chandra.
Radio VLBI, capable of achieving mas-scale resolution, is typically the only method by which dual AGNs with separations on the order of a few hundred parsecs ($\lesssim300$\,pc) can be resolved, and it is the only method by which gravitationally bound binary AGNs with separations $\lesssim30$\,pc can be unambiguously spatially resolved and confirmed at \cite[e.g.,][]{rodriguez2006}. However, while radio VLBI is a powerful technique for detecting incredibly close pairs of AGNs, the majority of AGNs do not exhibit significant radio emission \cite[e.g.,][]{padovani2017}, whereas X-ray emission is a ubiquitous feature of AGNs and probes the immediate vicinity of the central engine \cite[e.g.,][]{brandt2015}; mas-scale resolution in the X-rays could ultimately become a far more powerful tool than the VLA and the VLBA for probing galaxy mergers, sub-kpc dual AGNs, and pc-scale binary AGNs.  

\begin{figure*}
    \centering
    \includegraphics[width=0.9\linewidth]{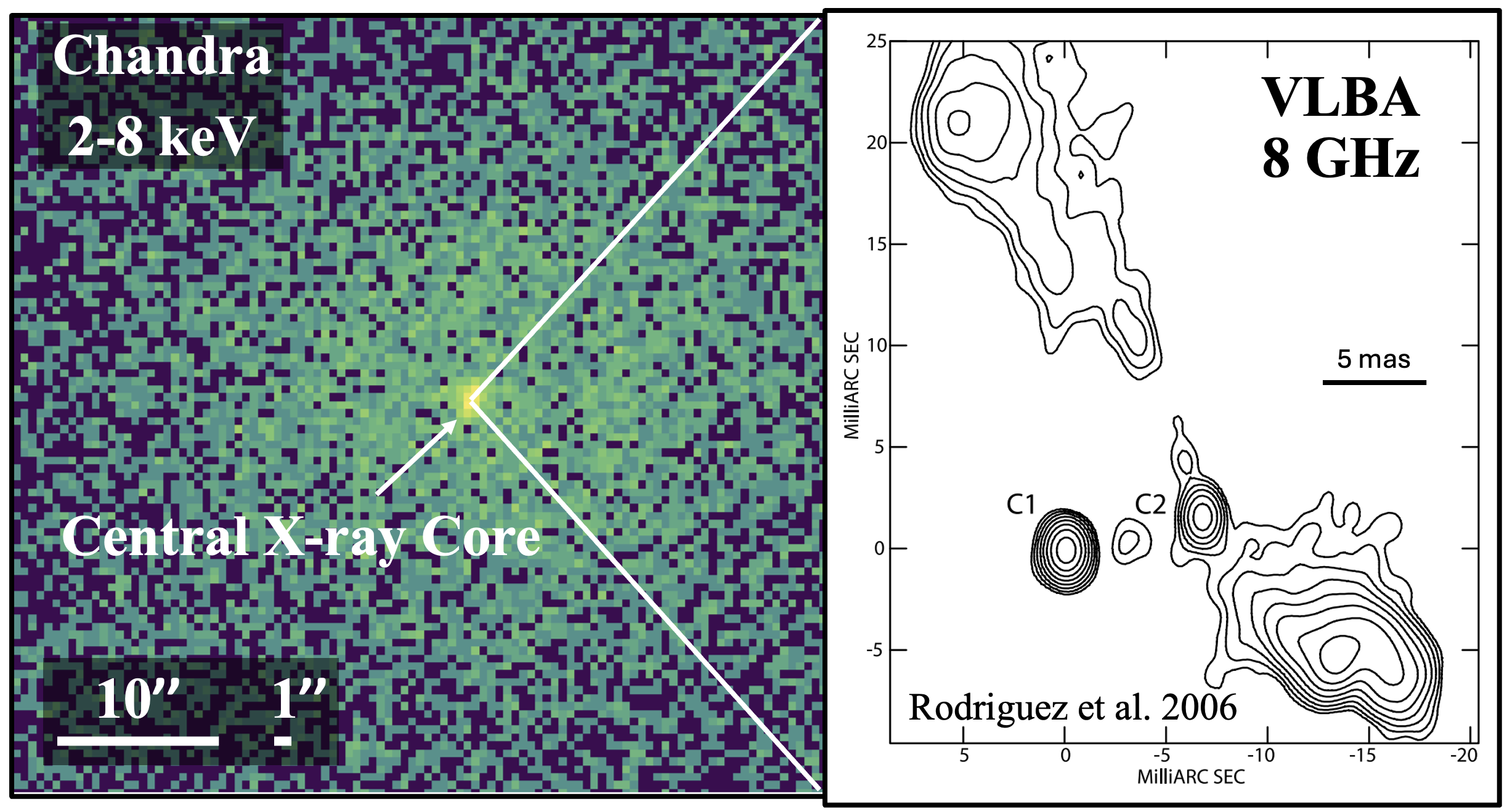}
    \caption{\textbf{Chandra X-ray and Very Long Baseline Array (VLBA) radio imaging of of CSO 0402+379, the only known binary AGN, with a separation of 7\,pc.} Left panel: Chandra $2-8$ keV imaging of CSO 0402+379, showing a bright $\sim$1'' central core and significant extended emission. The scale bars in the bottom left corner indicate angular scales of 10'' and 1''. Right panel: VLBA 8 GHz imaging from \cite{rodriguez2006} (reprinted with permission) showing two bright radio cores (C1 and C2) and extended emission associated with a jet emanating from C2; multi-frequency VLBA imaging confirmed C1 and C2 to be compact, flat-spectrum radio sources, consistent with AGNs. The scale bar on the right indicates an angular scale of 5 milliarcseconds. The angular resolution of Chandra is still $\sim2-3$ orders of magnitude too large to resolve binaries like CSO 0402+379; only mas-scale X-ray imaging missions can achieve the angular resolutions needed to spatially resolve binary AGNs. }
    \label{fig:binaryexample}
\end{figure*}

With 50\,mas resolution (and sufficient effective area), sub-kpc dual AGNs could be resolved down to separations of $\sim50$\,pc at $z=0.05$, $\sim90$\,pc at $z=0.1$, $\sim300$\,pc at $z=0.5$, and $\sim400$\,pc at $z=1.0$. In fact, in the very nearby universe ($z\sim0.02$) even bound binaries could be spatially resolved down to separations of $\sim10$\,pc with 50\,mas imaging. At higher redshifts, this sensitive 50-mas-scale imaging would unlock the ability to resolve kpc-scale and sub-kpc ($\gtrsim\,400$\,pc) dual AGNs, enabling the first spatially-resolved X-ray investigations of dual AGNs at cosmic noon and beyond, such as those being discovered now with JWST \cite[e.g.,][]{ishikawa2025,li2024}. $\sim1\,$mas resolution in concert with sufficient sensitivity, on the other hand, would provide the capability to resolve sub-pc binaries at $z\lesssim0.05$ and pc-scale binaries at higher redshift (Figure~\ref{Fig:resolution}). As a prime example, the only known binary AGN, CSO 0402+379 \citep[$z=0.05$, separation 7\,pc, Figure~\ref{fig:binaryexample},][]{rodriguez2006}, which was originally detected using radio VLBI, could be resolved in mas-scale X-ray imaging, enabling direct line-of-site obscuration constraints as well as a view of circumnuclear star formation, extended X-ray emission nebulae, and/or hot gas in the vicinity of the black holes. The fact that both nuclei are active (i.e., radio-bright) suggests ongoing accretion and implies dissipation of the angular momentum of the system. CSO 0402+379 was also detected in the \textit{ROSAT} All-Sky Survey and in follow-up \textit{Chandra} imaging \citep{romani2014}, making it a perfect target for high resolution X-ray imaging (see Figure~\ref{fig:binaryexample}). 

Sensitive, $\mu\rm{as}$ resolution would be even more revolutionary: it would open up the possibly of detecting milliparsec (mpc) scale binaries throughout the universe, including sub-mpc ($<1\times10^{-3}$\,pc) binaries at $z\lesssim0.05$). $\mu\rm{as}$ X-ray observations would have a dramatic impact on our understanding of binary accretion activity, accretion disk structure, and angular momentum losses at such separations, which would be impossible at lower angular resolutions. Furthermore, as mpc and sub-mpc binaries fall within the continuous gravitational wave regime \cite[e.g.][]{burke-spolaor2018,agazie2023b}, $\mu\rm{as}$ X-ray imaging and pulsar timing arrays (PTAs, \cite{agazie2023,agazie2023b}) or space-based gravitational wave detectors, such as the Laser Interferometer and Space Antenna (\textit{LISA} \cite{Amaro-Seoane2017}), could offer an unparalleled synergy capable of resolving distinct PTA sources and understanding the interplay between accretion activity, SMBH growth, and gravitational wave emission in these inspiralling binary systems. Whether through mas- or $\mu\rm{as}$-scale imaging, obtaining a larger X-ray sample of sub-kpc dual AGNs and binary AGNs would address key questions related to the merger-driven accretion activity, SMBH growth, and the overall binary population densities and sub-kpc SMBH pairing rates in merging/post-merger galaxies, all of which would transform our understanding of hierarchical growth in the universe. 

In addition to resolving distinct AGNs, ultra
high resolution X-ray imaging would also spatially resolve X-ray contributions from any surrounding hot gas and/or star formation driven-emission in these systems. Differentiating between AGN-driven and star formation-driven X-ray emission has been a persistent issue in the field of dual AGNs, especially when the observed X-ray luminosities of the dual AGNs are heavily suppressed due to high column densities and the AGN luminosities fall within the range expected for star forming galaxies ($L_{2-10\,\rm{keV}}<10^{42}$~erg~s$^{-1}$). Resolving the AGN and star formation emission separately would be one clear way of breaking this degeneracy, as the spectroscopic parameters and/or luminosity of the AGN emission could be distinctly discerned from star forming processes. Disentangling AGN and star forming emission is important, not only for resolving the central engines in these systems, but also for constraining (1) the star formation rates in these systems and, by extension, (2) the X-ray binary luminosity functions in these systems, and (3) the potential impact of AGN activity on star formation and galactic structure, among other questions.  

\subsection{Stellar science (an Astro2020 Habitable Worlds objective)}
\label{subsec:sciencecaseStellar}

High-resolution X-ray imaging would be pivotal for probing the most energetic phenomena associated with stars, including their dynamic outer atmospheres and interactions with orbiting planets. Pushing these observations to mas and $\mu$as angular resolutions would represent a transformative leap in observational capability, allowing us to resolve fine structures and dynamic processes in stellar environments that are currently blurred or entirely invisible.

Such a capability directly addresses key priorities outlined in Astro2020, which highlights high-energy stellar radiation as ``a risk to, and potentially a catalyst for, life," and emphasizes the need to place exoplanet observations in the broader context of stellar environments \citep{Astro2020Decadal}. Specifically, Appendix E poses critical questions about how host stars influence planetary atmospheric evolution (Appendix~E, Question~2d; E-Q2d) and how habitability varies with stellar type and evolutionary state (Appendix~E, Questions~3b–3d; E-Q3b–d). High-resolution X-ray imaging is uniquely positioned to answer these questions by directly tracing the magnetic energy release, flares, and winds that drive atmospheric escape, photochemistry, and long-term climate evolution.

With sub-mas resolution, we can conduct detailed studies of stellar flare morphologies, map magnetic field structures in stellar coronae, and observe real-time star–planet interactions. These observations will provide unprecedented insights into stellar magnetic activity and atmospheric escape processes—the fundamental mechanisms that determine whether worlds around other stars can support life as we know it.

\subsubsection{Stellar activity and its effects} 

\begin{figure}[!ht]
\centering
\small
\includegraphics[width=0.85\textwidth,angle=0,trim=0.5cm 1.5cm 0.5cm 1.0cm]{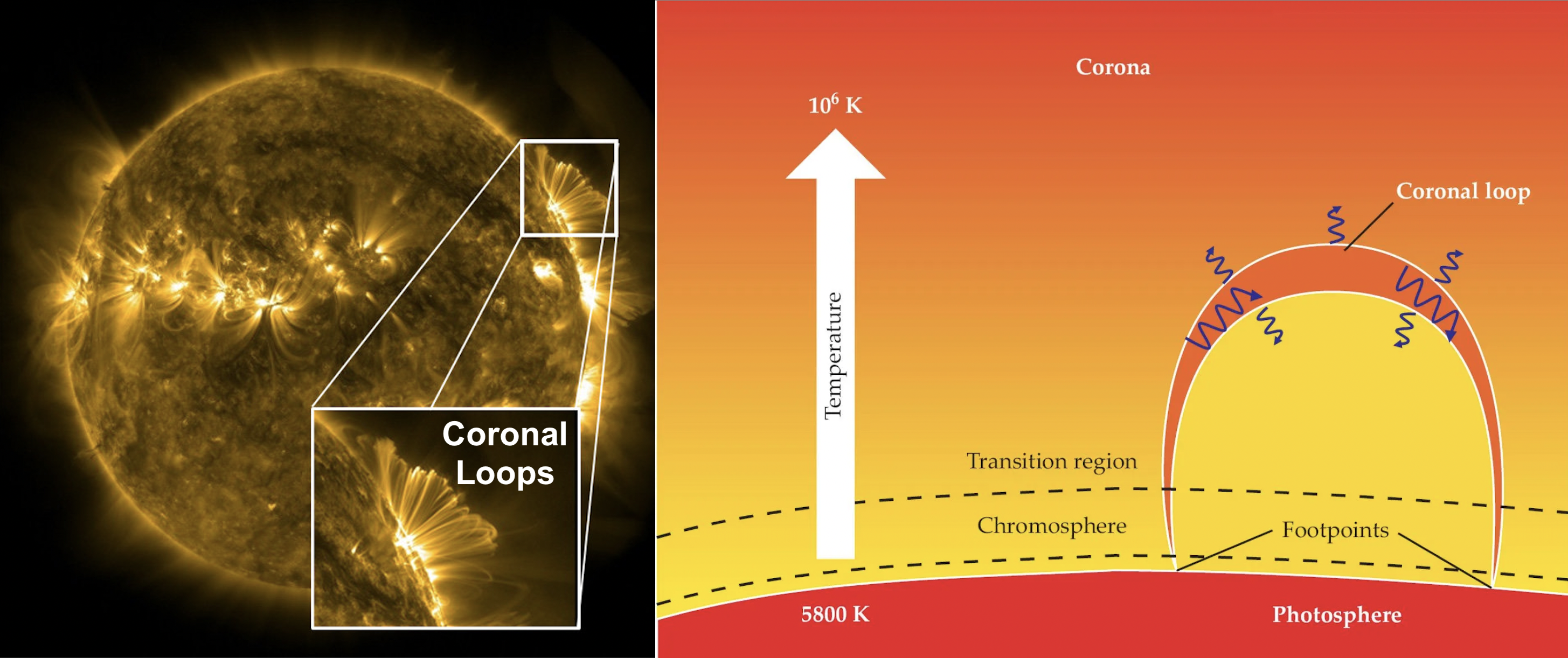}
\caption{
\textbf{An example of a coronal loop.}\
Left: An example solar image with a close up of a coronal loop region \citep{Moradi2021}. Right: An illustration of a coronal loop.  Coronal loops are anchored on both ends at footpoints in the photosphere; they project into the chromosphere and transition region and extend high into the corona. The wavy lines indicate how magnetohydrodynamic waves and heat propagate through one such loop. These propagating magnetohydrodynamic waves in loops could be possible explanations for coronal heating \citep{Sigalotti2023}.  Image credit: (left) NASA Solar Dynamic Observatory (right) L. Sigalotti, reprinted with permission.}
\label{fig:coronal_loops}
\end{figure}

Stellar coronae exhibit complex magnetic structures that fundamentally control stellar activity and its effects on their space weather. The arch-like coronal structures composed of relatively dense plasma confined by magnetic flux tubes are known as magnetic loops (see Figure~\ref{fig:coronal_loops}). These stellar loop structure plays a crucial role in our understanding of magnetic dynamo \citep{Reale-04-A+A-2, Reale-07-A+A-2, Karmakar-16-MNRAS-9,Karmakar-17-ApJ-5}. 
However, with current technology, these loop structures have never been spatially resolved. 

Sub-mas resolution observations would enable us to directly measure the sizes of nanoflares thought to produce bright coronal loops \citep{Pallavicini-90-A+A-6, Crespo-Chacon-07-A+A}. The capability of directly imaging the fine structures would enable us to measure their widths, temperatures, and densities \citep{Favata-00-A+A-3, Osten-10-ApJ-5, KarmakarS-22-MNRAS-1,KarmakarS-23-MNRAS-2}. 
These measurements would provide critical observational evidence to either support or refute the ``stranded loop" hypothesis—a key component in solving the long-standing coronal heating problem, which seeks to explain how stellar coronae reach temperatures of millions of degrees despite being farther from the stellar surface than the much cooler photosphere. Resolving this fundamental question in stellar physics would transform our ability to predict stellar radiation environments and assess atmospheric retention around exoplanets across diverse stellar types.

\vspace{-0.5ex}

A key advance enabled by high-resolution X-ray imaging would be the ability to directly compare coronal structures with Zeeman Doppler Imaging (ZDI) maps, which trace photospheric magnetic field concentrations. Such comparisons would provide crucial insight into the spatial extent and complexity of stellar active regions, testing whether features seen in ZDI maps represent large, coherent structures or the superposition of many smaller unresolved regions. This synergy between photospheric magnetic topology and coronal morphology would be further enriched by Doppler-resolved spectroscopy. A high-resolution capability would enable the mapping of Doppler shifts in X-ray emission lines along coronal loops with high spatial accuracy, yielding detailed information about velocity fields and their evolution. These measurements are vital for constraining models of loop dynamics and understanding the mechanisms responsible for heating plasma to coronal temperatures \citep{ArgiroffiC-19-NatAs-1}.

\begin{figure}[!ht]
\centering
\small
\includegraphics[width=0.78\textwidth,angle=0,trim=0.5cm 1.5cm 0.5cm 1cm]{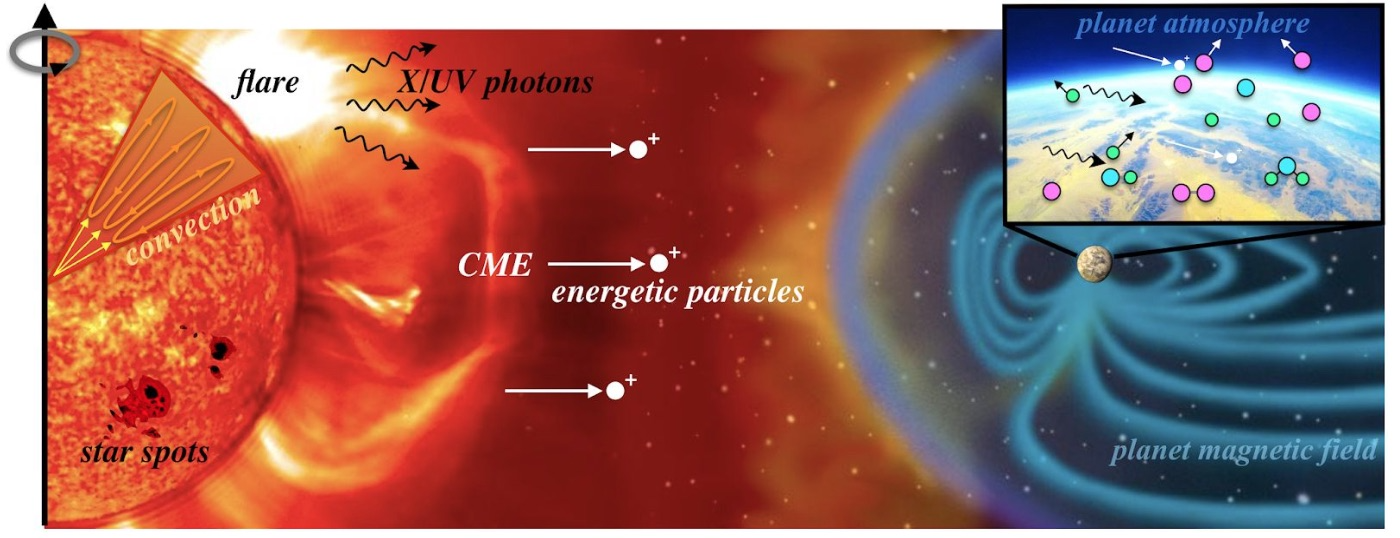}
\caption{
\textbf{High-resolution X-ray images would help assess extreme conditions on exoplanets orbiting solar-like stars.}\
Rotation and convection drive flares in low-mass stars. High-energy photons and energetic particles generated by flares and CMEs interact with planetary atmospheres (upper right inset) and lead to photochemistry, atmospheric loss, and impacts on habitability. Credit: NASA/GSFC Scientific Visualization Studio (SVS 30481); graphic design by S. Bensusen and S. Hill; adaptation by unknown.}
\label{fig:star_planet_interaction}
\end{figure}

Stellar space weather, characterized by energetic phenomena such as flares and coronal mass ejections (CMEs), plays a significant role in the evolution of stars and their planetary systems.
Stellar flares, which are sudden, intense releases of magnetic energy through reconnection, emit radiation across the entire electromagnetic spectrum (gamma-ray to radio) and/or kinetic energy in the form of CMEs. Both phenomena significantly affect planetary atmospheres by altering atmospheric chemistry, radiation levels, and the potential for liquid water on exoplanets \citep[][see Figure~\ref{fig:star_planet_interaction}]{Papaioannou2016,Guido2018}.

One of the key challenges in understanding these effects is resolving the locations of flare-producing active regions, especially when CMEs are involved. Although signatures of type-IV radio bursts have been linked to CMEs in a few M dwarfs, such as in the Sun, the spatial resolution needed to fully understand flare dynamics is currently lacking \citep{Odert2020_CMEs,AlvaradoGomez2022_CMEs_exoplanets,Leitzinger2024_CME_detection}. In a few M-dwarf systems, however, signatures of type-IV bursts have been found \citep{Zic_2020,atul24_ADLeotypeIV}, of which more than $\gtrsim$75\% are associated with fast CMEs on the Sun \citep{Kumari2022_solar_radio_burst,Mohan2024_solar_radio_bursts}. Spatial resolution is critical for tracing active regions and their evolution from the pre-flare to post-flare phases in X-ray bands, and for identifying local X-ray dimming signatures that may be associated with massive outflows. The locations of these active regions can indicate the direction of CME ejecta and their potential impacts on nearby planets. 
CMEs associated with major flares can persist from minutes to hours, as shown by solar and stellar observations. High-cadence image-plane tracking is therefore essential for determining the CME potential of large flares and for assessing their impacts on exoplanetary space weather.

\vspace{-0.5ex}

Coronal dimmings, as seen in X-ray and extreme-UV (EUV) emissions \citep{Veronig2021_xray_dimmings}, provide insights into the propagation direction \citep{Jain_2024_CME_direction} of CMEs and help distinguish between flares associated with mass ejections and those that are not \citep{Harrison2003_xray_dimmings,Moschou2017_CME}, with implications for atmospheric escape and exoplanet habitability \citep{Veronig2021_xray_dimmings}. Studies using Fe XII and Fe XXI emission lines from stars like $\epsilon$ Eridani further link coronal emission drops to CME mass \citep{Parke_Loyd_2022_CMEs}. A mas--$\mu\rm{as}$ resolution X-ray mission with high sensitivity will enable the detection of these dimmings from distant Sun-like stars, improving our understanding of stellar activity and its impact on exoplanetary environments.

High-resolution imaging of stellar coronae would provide direct constraints on how flares affect exoplanetary systems and habitability. Habitable-zone targets include Sun-like G and K dwarfs, as well as lower-mass M dwarfs \citep{Tarter2007,Kopparapu2013,Gordon2017}. High-resolution X-ray observations of these stars would reveal the extreme radiation and particle environments around these systems, clarifying how stellar activity drives atmospheric loss and space weather.

Planets in habitable zones around magnetically active stars may experience atmospheric erosion due to CMEs, exposing their surfaces to extreme flare activity and altering the potential for habitability \citep{Kay2016}. Solar research has shown there is a strong dependence on the latitude of the corona and the frequency of occurrence of CMEs \citep{Kane2008, Lin2022}. High-resolution X-ray images enable us to image the coronae of stars where the latitudinal distribution of CMEs could determine the frequency with which planets in the habitable zone are impacted. For example, Alpha Centauri A, a G2V solar twin with an angular diameter greater than 1~mas, provides a unique opportunity to study solar-like coronal activity and magnetic environments on a spatially resolved scale-even in the absence of a confirmed planet. In contrast, Proxima Centauri, an older but still active M dwarf with a known planet, offers a compelling case for studying the flare-driven atmospheric evolution of habitable-zone planets around low-mass stars. Together, these systems exemplify how high-resolution X-ray imaging can advance our understanding of stellar activity and its impact on exoplanetary atmospheres.

X-ray emission from stars in the cool half of the main sequence is generally attributed to the presence of a magnetic corona that contains plasma at temperatures exceeding $10^{6}$~K \citep{Gudel2004}. Coronae are ubiquitous among these stars, yet many of their fundamental magnetic field mechanisms are not fully understood and likely involve a combination of acoustic waves and magnetic reconnection events \citep{Stepien1989}. Spatially resolved coronal imaging provides valuable insight into the dynamic behavior and spatial distribution of energy release during intense flare events. Currently, the Sun is the only star for which we possess a spatially resolved coronal image. Achieving higher spatial resolution in X-ray observations is essential for extending these insights across stellar masses and evolutionary stages, and for determining how coronal properties vary. 

\subsubsection{Exoplanetary environment and habitability}%

The ability to resolve stellar coronae and flare origins with sub-mas resolution is key to advancing our understanding of exoplanetary habitability. Typical X-ray luminosities for cool dwarfs from spectral types A through M span from log$_{10}$ L$_{\text{X}}$ $\sim$ 26--30 erg s$^{-1}$ \citep{Freund_2018,Magaudda_2020,Mohan_2021_EMISSA}. For example, the M3.5V flare star AD Leo, located at $\sim$5~pc, has a mean coronal X-ray flux of $\sim$ 10$^{-11}$--$10^{-12}$~erg~s$^{-1}$~cm$^{-2}$ \citep{vandenBesselaar2003_ADLeo_XMM_Chandra}, which can increase by an order of magnitude or more during strong flares.

To study both quiescent and flaring activity, an instrument should achieve a sensitivity of $\sim 10^{-12}$--$10^{-13}$~erg~s$^{-1}$~cm$^{-2}$, on timescales averaging on the order of $\sim~<$2 hr. This sensitivity would enable sub-minute or even sub-second imaging of flare evolution, offering unprecedented details of stellar flares and their effects on exoplanets.
For nearby key targets like Proxima Centauri (1.3~pc), with known planets \citep{Faria_2022_proxcen_planet_candidate}, a sensitivity of $10^{-9}$–$10^{-11}$ erg s$^{-1}$ cm$^{-2}$ is required to detect X-ray fluxes. For a significant ($\sim$5$\sigma$) detection, the sensitivity needs to be five times better. Achieving $\mu$- or sub-mas resolution will allow the spatial resolution of flares on low-mass stars, which often remain unresolved point sources, limiting our understanding of where flares originate on the stellar surface.

Sub-second particle acceleration signatures in radio and X-ray imaging of solar active regions highlight the importance of coordinated observations to track active region latitudes and evolution, distinguishing thermal and non-thermal counterparts in the low corona and chromosphere \cite[e.g.][]{Morosan14_millisecLOFARimging,Mohan17_SubsecSolarvariability,Chen19_VLAMillisecImging,Mohan_2019,Mohan23_solar-stellarSnapshotimaging,Juan23_ALMASec-scaleChromosphere,skokic23_all_Mmflares_sun_multibandstudy}. The ability to resolve stellar flares and their latitudinal distribution would provide critical insights into the dynamics of flare-producing active regions. Such detailed observations will help us understand how stellar activity influences exoplanetary atmospheres, including the erosion of atmospheres due to CMEs, as well as the potential for life on planets exposed to these extreme environments. High resolution X-ray imaging could also enhance multiwavelength coordination with observatories like the upcoming Square Kilometer Array (SKA) \citep{Braun_2019_arXiv_SKA1}, which will provide dynamic radio spectra from $\sim$50 MHz to 15 GHz. This synergy will improve our ability to study stellar space weather and its impact on exoplanetary environments.

High-energy stellar radiation plays a dual role: it can drive atmospheric mass loss and challenge habitability, but it also plays a critical role in prebiotic chemistry \cite[e.g.][]{2007AsBio...7..185L,airapetian17,Maliniemi22_SEP_depleteOzoneEarth,airapetian21_adv_n_disadv_SEPsforlife}, which is essential for life formation. By extracting light curves from high-resolution X-ray images, we could analyze flare morphology to distinguish superimposed flares and characterize flare frequency distributions (FFDs) of planet-host stars, and we could assess the impact of stellar activity on exoplanet atmospheres. This would help clarify the extent to which high-energy radiation from stellar flares may influence abiogenesis, defined as the chemical pathways that generate life’s molecular precursors on early Earth-like exoplanets \citep{Rimmer2018}. 

\subsection{Activity in Young Stellar Objects (an Astro2020 Habitable Worlds objective):}

Young Stellar objects (YSOs) are active signposts of stellar evolution and are frequently strong sources of X-rays (see, e.g., \citet{1999ARA&A..37..363F}). These objects represent the next stages of the evolution of protostars, where angular momentum is being lost, and the circumstellar envelope and surrounding stellar disk are accreting material. YSOs are often accompanied by jet-like mass outflows extending perpendicular to the disk in a biconical geometry. However, it is unknown from where or how the jets are launched (e.g., Are these stellar winds or disk winds?). One proposed idea is that jets/winds are launched at the inner edge of the accretion disk and represent the centrifugal ejection of material from the disk \citep[see ][and references therein]{Ray_Ferreira2021_youngstar_jets}. In fact, jets may be key to the process of star formation and planet formation because removal of angular momentum from the protoplanetary disk may allow accretion to proceed and may even determine the initial conditions for planet formation. 

Observations reaching mas resolution in the mm regime with facilities such as ALMA have revealed the structure of the launching region within $\sim$50 AU and the evolving kinematics of the gas. X-ray emission from these jets seems limited to within scales of order 1000 AU, which is only a few as at distances beyond 100 pc \citep{schwartz2010}. This small size scale requires an X-ray telescope with 0.1-0.001 as resolution to extract the astrophysics of jet formation and jet collimation taking place near the proto-stellar object. Several of these systems show extended X-ray emission at $\sim0.5$ as scales with Chandra.

The nearby YSO DG Tau offers a tantalizing clue to the physics of the launching region on small spatial scales. The soft and hard X-ray components observed with the intrinsic energy resolution of the Chandra ACIS-S are separated spatially by approximately 0.2 as \citep{schneiderschmitt2008}, which is a physical distance of 45 AU from the central source, assuming a distance of 140 pc. These authors demonstrate that this is a true physical offset, and the peak of the soft X-rays also coincides with peaks in optical emission. The coincidence implies the same physical origin for the optical and soft X-ray emission. Thus, DG Tau would be a key protostellar target for ultra high-resolution narrow band (soft and hard) X-ray imaging.

The accreting T Tauri star RY Tau has a significant optical jet and counter jet \citep{2019A&A...628A..68G}. It also displays an X-ray jet, but it is not clear what produces the emission \citep{2011ApJ...737...19S}. The temperatures for a shock-heated jet that are predicted on the basis of optical estimates are too low to explain the X-ray structure. Possibilities are high-speed material that has escaped detection at longer wavelengths, the result of magnetic heating, or hot plasmoids being ejected at high speeds. With a capability for mas resolution at X-ray energies, we could precisely learn the origins of the X-rays - whether they are related to the process that produces the optical jet or to the accretion process.

\section{Current or potential high-resolution imaging technologies}

There are several early-stage efforts ongoing to surpass technology development beyond the state-of-the-art of $\sim0.5-1$ as enabled with traditional X-ray mirrors \citep{2018SPIE10699E..0OZ}. In the sub-as regime, work has been done to improve the resolving power of X-ray mirrors by improving polishing, mounting stability, and alignment. Such work for developing new, high-precision X-ray telescope mirror technology would be critically necessary for missions such as a future flagship with sub-as HPD angular resolution goals (Table \ref{tab:tech}). Alternatively, there is current development on thin silicon optics with femtosecond laser technologies to correct for mirror figure errors and potentially increase resolution \citep{Zuo:22}. Mirrors are not the only option. There has also been recently funded work to fabricate diffractive X-ray lenses to achieve angular resolutions better than $0.1''$. Phase fresnel lenses can provide resolutions of $\sim20$ mas (\citet{2020arXiv200812810K}, Table \ref{tab:tech}).

There has similarly been technology development performed on the potential for X-ray interferometry to improve spatial resolution. The first major technical hurdle for X-ray interferometry with a distributed aperture was surpassed a quarter century ago when X-ray fringes were demonstrated in the lab \citep{cash2000}. Soon thereafter, a NASA Innovative Advanced Concepts (NIAC) study for the Micro-arcsecond X-ray Imaging Mission (MAXIM) was led by Webster Cash in the early 2000s, and tests were performed in the lab \citep{gendreau2004}. \citet{cash2003a,cash2003b} outlines a mission architecture for a diffraction limited structure with a 200 m baseline and 450 km focal length. MAXIM was proposed to NASA as a Vision mission in 2003 with a pathfinder version, but this ultimately was not adopted and thus progress stalled. We note another architecture that has been proposed to achieve ultra high-resolution imaging, which is an Aragoscope\footnote{\href{https://www.nasa.gov/wp-content/uploads/2017/03/2014_phase_i_webster_cash_the_aragoscope.pdf}{https://www.nasa.gov/wp-content/uploads/2017/03/2014\_phase\_i\_webster\_cash
\_the\_aragoscope.pdf}}. This is a telescope with an occulting disk as its primary optic, with light that is diffracted from a distant source around its edge and then focused along the optical axis. To image the event horizons of SMBHs in AGNs would require a primary Aragoscope disk of about one kilometer in diameter. 

 Since those earlier interferometry studies, a ``telephoto'' interferometry design has been created to potentially fit within a single spacecraft \citep{willingale2004}. A testbed has been assessed \citep{denhartog2020} and current best assessments for future interferometry flight technology can be found in discussions by \citep{uttley2020}.
The European Space Agency solicited ideas for its Voyage 2050, and \citet{uttley2021b} proposed an X-ray imaging interferometer with a compact optical design \citep{willingale2004}. A testbed was considered \citep{denhartog2020}; however, this mission was also not adopted. The status of ESA's plans for interferometry is discussed by \cite{Uttley2021}, and some technology development is ongoing (Table \ref{tab:tech}).

\section{Mission Concept: Accretion Explorer Interferometer} 
Our team is pursuing a NASA Innovative Advanced Concepts (NIAC) study to design a modernized X-ray interferometer.
\footnote{\href{https://www.nasa.gov/directorates/stmd/niac/beholding-black-hole-power-with-the-accretion-explorer-interferometer/}{https://www.nasa.gov/directorates/stmd/niac/beholding-black-hole-power-with-the-accretion-explorer-interferometer/}}
This would be a multiple spacecraft system that provides the architecture to align individual mirror pair baseline groupings provided by individual collector mirror spacecraft, with the pointing precision to achieve $\mu$as resolution. A mission design lab (MDL) architectural study was completed at Goddard Space Flight Center in 2025 and a report is being drafted for publication (Weaver et al., in prep). We have assessed various tall poles such as the pointing stability on target, which could be a primary obstacle to obtaining the long exposure times necessary for significant photon collection for many astrophysical sources. The visibility of interferometric fringes can be significantly reduced if the variations in optical path difference become large compared to the X-ray wavelength \citep{Uttley2021}, so stability on the order of nm is needed, possibly on the timescales of days, immune to spacecraft vibrations. 

Our X-ray interferometer would achieve $\mu$as resolution over scientifically motivated energy bands, based on the science investigations here and on input from the wider community. We are revisiting some aspects of the MAXIM study after many years of technology evolution in the field, including tremendous advancements in formation flying \citep[discussed by ][]{monnier2019}, and the pointing precision and stability demonstrated with JWST and the LISA pathfinder. LISA has, in fact, developed extremely precise measurements with a laser interferometer – measuring relative distance between two test mirrors with an accuracy of a few picometers per square root Hertz \citep{armano2021}, which seems to be adequate for the stability required to maintain X-ray interferometry baselines. Our Accretion Explorer architecture is substantially different from MAXIM, enabling larger effective areas per energy channel with a combination of large collecting mirror area, high combiner optic throughput and large-area CCD detectors. Our combiner optic under study would fly on the detector craft and would eliminate the long focus lengths of the MAXIM and Uttley designs, dramatically decreasing the distances between the formation-flying mirror to detector spacecraft that were previously required.

\begin{figure}
\centering
\includegraphics[width=0.85\linewidth]{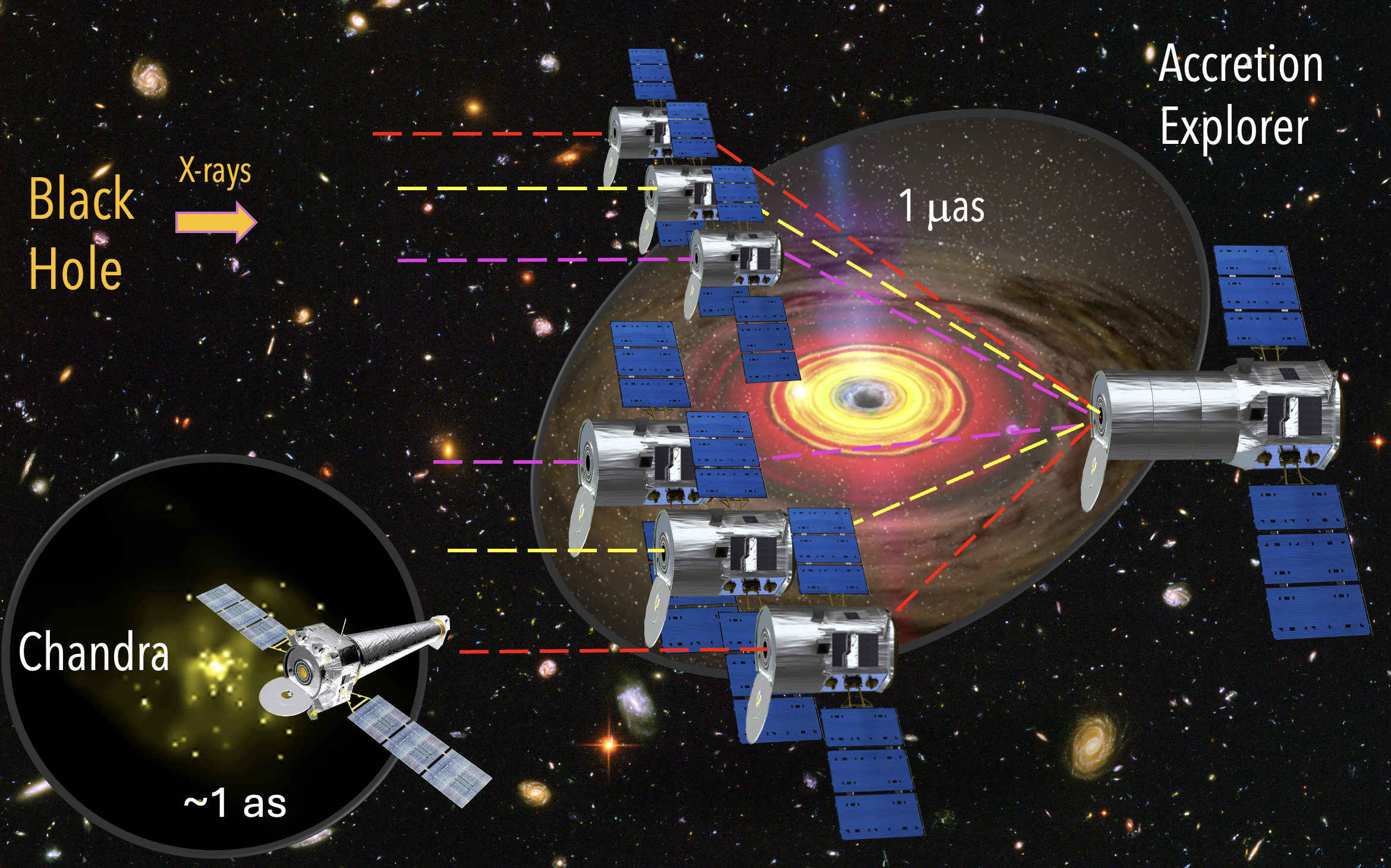}
\caption{
\textbf{Artist's Concept for the Accretion Explorer Interferometer design.} This figure shows a representative array design for a $\mu$as interferometer with three energy channels provided by six identical collector satellites. The baselines range from $\sim20$ m to $\sim100$ m and the distance between the collector satellites and detector satellite is $\sim2$ km. Credit: B. Griswold and K. Weaver.}
\label{fig:AEI}
\end{figure}

 For a defined energy range, a set of primary mirrors housed in a pair of collector spacecraft would collect X-ray photons and direct them via grazing angle reflection to a secondary optic housed in the detector spacecraft. This secondary optic would steer the X-rays to a common focal plane. We are studying initially three energy channels, based on mapping key emission lines, ($0.7-1.2$ keV), ($1.5-2.5$ keV), ($6-7$ keV), provided with three sets of collector mirror spacecraft with different baselines. If the array were operated so as to move the mirror craft to expand or contract baselines, these energy ranges can shift.

Our MDL study looked at the top level mission architecture to determine if such a system is feasible. It specifically addressed formation flying requirements, launch vehicles and launch configurations, mass, volume, and power requirements, satellite control and maneuverability, and communications and data rates. We have also examined the most suitable orbit that maximizes the stability for pointing on target. Some decisions have been made in terms of trade-offs.  We examined the small $\sim20$ baselines of the inner collector mirror pairs. Even though there are risks of small baselines \citep{monnier2019}, we prefer free flyers instead of tethering the spacecraft, with the operational risks of tethers being slightly higher. We were able to obtain the required pointing precision and extreme stability on target by utilizing several pairs of LISA telescopes; however, the extreme stability of the array requires a large amount of fuel for standard satellite operations. This can limit mission life, but there are other operational considerations that require further study to mitigate a short (few years) lifetime constraint.

Overall, our study found no show stoppers for a mission architecture that could provide narrow-band $\mu$as X-ray imaging, presuming the individual optical elements can reach high enough technology readiness levels (TRLs). The interferometer will require several X-ray finder scopes to lock on target and step up in resolving power from the as to $\mu$as scale. These telescopes will likely all have a different technology heritage. To anticipate the complexity of creating a future integrated spacecraft, we have designed a modular system with locations to plug in these different telescopes within the spacecraft volume. This modular design provides hope for a future feasible flagship or probe class mission coming together quickly once the basic technology development has reached the needed TRL. 

\section{Summary}

 We have presented the broad science case for obtaining mas to $\mu$as astronomical imaging resolution in the soft to medium-energy X-ray band ($\sim0.5$ to $\sim8$ keV). Because X-ray imaging capabilities have fallen so far behind observations at longer wavelengths, we advocate now for decisive advances in X-ray imaging if we can hope to achieve these Astro2020 science goals within the next two decades. 
 
 We have shown that merely $0.1''$ resolution would significantly advance our understanding of AGN feedback. A capability for mas resolution could provide a wealth of information on BH winds and the AGN torus. But $\mu$as resolution is required to reach scales on the sizes of the X-ray coronae and accretion disks. Studies of jet launching regions also require the highest resolutions. Each step up in resolving power will revolutionize studies of XRB populations and reach ever higher redshifts. The same is true for dual AGNs, while sensitive, $\mu\rm{as}$ resolution would be revolutionary in opening up the possibility of detecting milliparsec scale SMBH binaries throughout the universe.

 High-resolution X-ray imaging would be pivotal for probing the most energetic phenomena associated with stars, including their dynamic outer atmospheres and interactions with orbiting planets. Pushing these observations to mas and $\mu$as angular resolutions  resolve fine structures and dynamic processes in stellar environments. Sub-mas resolution would allow detailed studies of stellar flare morphologies, map magnetic field structures in stellar coronae, and provide unprecedented insights into the atmospheric escape processes in exoplanets. Habitable-zone targets for interferometry would include Sun-like G and K dwarfs, as well as lower-mass M dwarfs. 
 
 We have also provided a top level summary of current technologies that are under study to advance these goals. In addition to mentioning developments in high quality mirrors and lenses to improve more traditional methods from the current $\sim0.5''$ to the mas scale, we discuss the Accretion Explorer, a mission architecture for a dispersed aperture X-ray interferometer. We believe this architecture study will be helpful to create a path for these technologies to develop and move forward with an already feasible mission design framework in place. This is an example of a modernized concept that could succeed for the purpose of enabling ultra high angular resolution.

\section{Figures}
Frontiers requires figures to be submitted individually, in the same order as they are referred to in the manuscript. 
\section{Tables}

\begin{table}[h!]
    \centering
    \caption{Technology and Architecture Studies: This table lists current or very recent technology and architecture studies that have promise for advancing X-ray imaging capability. Column (1) lists the approximate resolution range being targeted. Column (2) is the specific technique being developed. Column (3) lists the lead organization, although many of these efforts have wide collaboration.  MIT is the Massachusettes Institute of Technology. GSFC is NASA's Goddard  Space Flight Center. ESA is the European Space Agency. Column (4) lists the most recent reference for the work, when available.} 
    
    \begin{tabular}{|| l | c | c | c ||}
    \hline
    \multicolumn{4}{|c|}{Technology and Architecture Studies} \\
    \hline\hline
    Resolution & Technique & Lead & Ref. \\
    (1) & (2) & (3) & (4) \\
    \hline\hline
$\leq 0.1$ arcsec & polished silicon mirrors & MIT & https://techport.nasa.gov/projects/96334   \\
$\leq 0.1$ arcsec &  multilayer-coated mirrors & MIT  &  Herman Marshall, private communication  \\
$\leq 0.1$ arcsec & diffractive X-ray lenses &  GSFC  & https://techport.nasa.gov/projects/146930  \\
Milli-arcsec & interferometry        & GSFC     & https://techport.nasa.gov/view/117143    \\
20-50 mas & phase fresnel lenses    &   GSFC     & \citet{2020arXiv200812810K}   \\
Milli-micro arcsec & interferometry  & ESA      &   \citet{uttley2021b}  \\
Micro-arcsec & interferometry       & GSFC      & This article \\
\hline
    \end{tabular}
    \label{tab:tech}
\end{table}

\section*{Conflict of Interest Statement}

The authors declare that the research was conducted in the absence of any commercial or financial relationships that could be construed as a potential conflict of interest.

\section*{Author Contributions}

KAW provided the initial draft of the manuscript, conceived the work, managed top-level organization, provided figures, served as executive editor, and collaborated in aims and goals. 
JMC provided support to the drafting of the AGN science sections and assisted KAW in organization and editing. 
RWP led the drafting in the XRB Science Section and the Dual AGN and Binary AGN Science Section, including the development of the science cases and supporting figures contained therein as well as providing revisions in the final version. 
MM aided in drafting the dual AGN, XRB and relativistic jet sections and in the literature review.
LDV led the drafting of the Stellar Science section, contributed to the literature review and framing of the science case, and provided revisions to that section in the final manuscript.
RG aided in writing the relativistic jet section. 
TM aided the support to the drafting to the Stellar Science section.
KM aided in drafting the Stellar Science and Young Stellar Objects sections and literature review. 
MS aided in drafting the BH winds and feedback and relativistic jet sections.
EL aided in drafting the SMBH accretion and AGN corona section.
JDM provided the \textsc{RefleX} simulations and assisted in drafting the AGN science sections. 
IC contributed to the AGN sections.
KEW contributed to the sections on feedback/outflows and XRB populations. 
EK contributed to the AGN sections.
AM contributed to the stellar science sections.
SJ contributed to the stellar science section.
AH contributed to the XRB section.
HO contributed to the stellar science section.
CR, KD, LV, TM, RP and BC provided revision suggestions for the final draft of the manuscript.

\section*{Funding}

This material is based on work supported by the NASA Innovative Advanced Concepts (NIAC) program and by NASA under award number 80GSFC24M0006. 

\section*{Acknowledgments}

[J.D.M.'s] research was supported by an appointment to the NASA Postdoctoral Program at the NASA Goddard Space Flight Center, administered by Oak Ridge Associated Universities under contract with NASA.

\bibliographystyle{Frontiers-Harvard} 
\bibliography{test}


\end{document}